%% file: thesis.tex
\pdfoutput=1
\documentclass
[
    openright,               
    cleardoublepage = empty, 
    fontsize = 12 pt,        
    american,                
    captions = tableheading, 
    numbers = noenddot,      
    footheight = 35 pt,      
]
{scrbook}

\input{core/preamble/document_settings} 

    \extraBorder{3 mm}

    \bindingCorrection{6 mm}

    \boldNumberSetsfalse 

    \myTitle{Behaviorally Correct Learning\newline from Informants}{Verhaltenskorrektes Lernen von Informanten}

    \myName{Niklas Mohrin}

    \myProgram{IT-Systems Engineering}

    \dateOfHandingIn{30. Juni 2022}

    \nameOfMySupervisor{Prof.\,Dr. Tobias Friedrich}

    \additionalExaminers{Dr. Timo Kötzing\newline Vanja Dosko\v{c}}

    \mySubject{We compare semantic restrictions in the context of behaviorally correct learning from informants.}

    \myKeywords{bachelor thesis | learning theory | inductive inference | language learning in the limit | semantic learning | behaviorally correct | informant | cautiousness | monotonicity | dual monotonicity | consistency}

\input{core/preamble/format}                        
\input{core/bibliography/bibliography_format}       
\input{packages_and_commands/standard_packages}     
\input{packages_and_commands/general_commands}      
\input{packages_and_commands/user-defined_commands} 

\begin{document}

    \frontmatter
    \include{core/title_page/title_page}

    \pagestyle{plain}

    \addchap{Abstract}
    \input{preface/abstract}

    \selectlanguage{ngerman}
    \addchap{Zusammenfassung}
    \input{preface/abstractGerman}
    \selectlanguage{american}


    \setuptoc{toc}{totoc}
    \tableofcontents

    \pagestyle{headings}
    \mainmatter

    \chapter{Introduction}
    \input{introduction/introduction}

    \chapter{Preliminaries}
    \input{introduction/preliminaries}

    \chapter{General observations}
    \input{core/plain}

    \chapter{Cautiousness}
    \input{core/cautiousness}

    \chapter{Monotonicity}
    \input{core/monotonicity}

    \chapter{Consistency}
    \input{core/consistency}



    \makeatletter
        \def\toclevel@chapter{-1}
        \def\toclevel@section{0}
    \makeatother

    \chapter{Further research}
    \input{conclusions/further_research}

    \pagestyle{plain}

    \renewcommand*{\bibfont}{\small}
    \printbibheading
    \addcontentsline{toc}{chapter}{Bibliography}
    \printbibliography[heading = none]


\end{document}

%% file: core/preamble/document_settings.tex
\newif\ifprintVersion   
\newif\ifprofessionalPrint 
\newif\iffancyTheorems  
\newif\ifboldNumberSets 
\newif\ifbachelorThesis 

\printVersionfalse
\professionalPrintfalse
\fancyTheoremstrue
\boldNumberSetstrue
\bachelorThesistrue

\newcommand*{\printTitle}{}
\newcommand*{\printGermanTitle}{}
\newcommand*{\myTitle}[2]{\renewcommand*{\printTitle}{#1}\renewcommand*{\printGermanTitle}{#2}}
\newcommand*{\printTitleBold}{\textbf{\printTitle}}

\newcommand*{\printAuthor}{}
\newcommand*{\myName}[1]{\renewcommand*{\printAuthor}{#1}}

\newcommand*{\printProgram}{}
\newcommand*{\myProgram}[1]{\renewcommand*{\printProgram}{#1}}

\newcommand*{\printDateReceived}{}
\newcommand*{\dateOfHandingIn}[1]{\renewcommand*{\printDateReceived}{#1}}

\newcommand*{\printSubject}{}
\newcommand*{\mySubject}[1]{\renewcommand*{\printSubject}{#1}}

\newcommand*{\printKeywords}{}
\newcommand*{\myKeywords}[1]{\renewcommand*{\printKeywords}{#1}}

\newcommand*{\printNameOfSupervisor}{}
\newcommand*{\nameOfMySupervisor}[1]{\renewcommand*{\printNameOfSupervisor}{#1}}

\newcommand*{\printAdditionalExaminers}{}
\newcommand*{\additionalExaminers}[1]{\renewcommand*{\printAdditionalExaminers}{#1}}

\newlength{\extraborderlength}
\newcommand*{\extraBorder}[1]{\setlength{\extraborderlength}{#1}}

\newlength{\mybindingcorrection}
\newcommand*{\bindingCorrection}[1]{\setlength{\mybindingcorrection}{#1}}

%% file: core/preamble/format.tex
\usepackage[utf8]{inputenc} 
\usepackage[T1]{fontenc}    
\usepackage
[
    ngerman,         
    main = american, 
]
{babel}                     

\input glyphtounicode
\usepackage{calc} 

\newlength{\myparindent}
\newlength{\myparskip}
\setfootnoterule{0 cm}

\deffootnote[1.2 em]{1.2 em}{0 em}{\makebox[1.4 em][l]{\textbf{\thefootnotemark}}}

\makeatletter%
    \@removefromreset{footnote}{chapter}%
\makeatother

\usepackage[dvipsnames]{xcolor} 

\definecolor{stroke1}{HTML}{2574A9} 

\colorlet{captionlabel}{black}
\colorlet{footerpagenr}{black}
\colorlet{footerchapter}{stroke1}
\colorlet{footerchaptername}{black}
\colorlet{footersection}{stroke1}
\colorlet{footersectionname}{black}
\colorlet{chapternumber}{stroke1}

\newlength{\mypaperwidth}
\newlength{\mypaperheight}
\newlength{\mybodywidth}
\newlength{\mybodyheight}
\newlength{\myoutermargin}
\ifprintVersion
    \ifprofessionalPrint
    \else
    \fi
\else
\fi

\newlength{\mytopmargin}
\ifprintVersion
    \ifprofessionalPrint
    \fi
\fi

\newlength{\myinnermargin}
\ifprintVersion
    \ifprofessionalPrint
    \fi
\fi

\newlength{\mybottommargin}
\ifprintVersion
    \ifprofessionalPrint
    \fi
\fi

\newcommand{\goldenratio}{1.618}

\newlength{\myheadsep} 
\ifprintVersion
    \ifprofessionalPrint
    \fi
\fi

\newlength{\myfootskip} 
\ifprintVersion
    \ifprofessionalPrint
    \fi
\fi

\newlength{\mymargininnersep} 
\newlength{\mymarginoutersep} 
\ifprintVersion
    \ifprofessionalPrint
    \fi
\fi

\newlength{\mymarginwidth} 
\newlength{\mymarginwidthwithinnersep} 
\usepackage
[
    \ifprintVersion
        \ifprofessionalPrint
            paperwidth = \mypaperwidth + 2\extraborderlength + \mybindingcorrection,
            paperheight = \mypaperheight + 2\extraborderlength,
        \else
            paperwidth = \mypaperwidth,
            paperheight = \mypaperheight,
        \fi
    \else
        paperwidth = \mypaperwidth,
        paperheight = \mypaperheight,
    \fi
    textwidth = \mybodywidth,
    textheight = \mybodyheight,
    outer = \myoutermargin,
    top = \mytopmargin,
    headsep = \myheadsep,
    footskip = \myfootskip,
    marginparsep = \mymargininnersep,
    marginparwidth = \mymarginwidth,
]
{geometry} 

\usepackage
[
]
{scrlayer-scrpage} 

\KOMAoptions
{%
    headwidth = \textwidth + \mymarginwidthwithinnersep,%
    footwidth = \myoutermargin : \textwidth,%
}

\automark[chapter]{chapter}
\automark*[section]{}

\lehead%
{%
    \begin{minipage}[b]{\mymarginwidth}%
        \small\raggedleft\normalfont\textsf{\textbf{\color{footerchapter}\chaptername\ \thechapter}}
    \end{minipage}
}
\cehead{\hspace*{\mymarginwidthwithinnersep}\parbox{\textwidth}{\raggedright\leftmark}}

\rohead%
{%
    \Ifstr{\rightmark}{\leftmark}%
    {%
        \begin{minipage}[b]{\mymarginwidth}%
            \small\raggedright\normalfont\textsf{\textbf{\color{footersection}Chapter\ \thechapter}}%
        \end{minipage}%
    }%
    {%
        \begin{minipage}[b]{\mymarginwidth}%
            \small\raggedright\normalfont\textsf{\textbf{\color{footersection}Section\ \thesection}}%
        \end{minipage}%
    }%
}
\cohead{\hspace*{-\mymarginwidthwithinnersep}\parbox{\textwidth}{\raggedleft\rightmark}}

\lefoot*%
{%
    \vspace*{1 ex}%
    {\color{stroke1}\rule{\myoutermargin - \mymargininnersep}{0.5 mm}}\\
    \begin{minipage}[b]{\myoutermargin - \mymargininnersep}%
        \raggedleft\normalfont\color{footerpagenr}\textbf{\thepage}%
    \end{minipage}%
}
\rofoot*%
{%
    {\color{stroke1}\rule{\myoutermargin - \mymargininnersep}{0.5 mm}}\\
    \begin{minipage}[b]{\myoutermargin - \mymargininnersep}%
        \raggedright\normalfont\color{footerpagenr}\textbf{\thepage}%
    \end{minipage}%
}

\usepackage{caption}
\usepackage{tikz} 

\newlength{\mytmpa}
\newlength{\mytmpb}

\renewcommand*{\partlineswithprefixformat}[3]%
{%
    #2
    \thispagestyle{empty}
    \setlength{\mytmpa}{0.618\mypaperwidth}%
    \setlength{\mytmpb}{0.382\mypaperheight}%
    \ifprintVersion
        \ifprofessionalPrint
            \setlength{\mytmpa}{0.618\mypaperwidth + \mybindingcorrection + \extraborderlength}%
            \setlength{\mytmpb}{0.382\mypaperheight + \extraborderlength}%
        \fi
    \fi
    \begin{tikzpicture}[overlay, remember picture]%
        \node [inner sep = 0, outer sep = 0, anchor = north] at (current page.north west)%
        {%
            \begin{tikzpicture}[overlay, remember picture]%
            \draw[color = stroke1, line width = 0.7 mm] (\mytmpa, 0) -- (\mytmpa, -\mytmpb);%
            \end{tikzpicture}%
        };%
        \node (align) [align = right, below = \mytmpb - 2 ex, inner sep = 0, outer sep = 0, anchor = north west] at (current page.north west)%
        {%
            \hspace{\mytmpa}\hspace{0.5 em}\partname\ \thepart\\[1 ex]
            \color{stroke1}#3%
        };%
    \end{tikzpicture}%
}
\RedeclareSectionCommand%
[%
    font = \normalfont\Huge\sffamily,
    prefixfont = \normalfont\Huge\sffamily,
]
{part}

\usepackage{etoolbox}

\newbool{chapterHasANumber}
\newbool{chapterHasAStar}
\renewcommand*{\chapterlinesformat}[3]%
{%
    \Ifnumbered{#1}{\setbool{chapterHasANumber}{true}}{\setbool{chapterHasANumber}{false}}%
    \Ifstr{#2}{}{\setbool{chapterHasAStar}{true}}{\setbool{chapterHasAStar}{false}}%
    \ifboolexpr{bool{chapterHasANumber} and not bool{chapterHasAStar}}%
    {%
        \begin{tikzpicture}[overlay, remember picture]%
            \node [right = \myinnermargin, below = \mytopmargin, inner sep = 0, outer sep = 0, anchor = north west] (numbernode) at (current page.north west)%
            {%
                \hspace{\myinnermargin}%
                \sffamily\fontsize{60}{60}\selectfont%
                \color{chapternumber}%
                \thechapter%
            };%
            \node [inner sep = 0, outer sep = 0, anchor = north west] at (numbernode.south west)%
            {%
                \begin{tikzpicture}[overlay, remember picture]%
                    \draw[color = stroke1, line width = 0.7 mm] (\myinnermargin, -1 ex) -- (\paperwidth, -1 ex);%
                \end{tikzpicture}%
            };%
            \node (align) [text width = \textwidth - 2 cm, align = right, right = \myinnermargin + \mybodywidth, inner sep = 0, outer sep = 0, anchor = east] at (numbernode.west)%
            {%
                #3%
            };%
        \end{tikzpicture}%
    }%
    {%
        \begin{tikzpicture}[overlay, remember picture]%
            \node [right = \myinnermargin, below = \mytopmargin, inner sep = 0, outer sep = 0, anchor = north west] (numbernode) at (current page.north west)%
            {%
                \hspace{\myinnermargin}%
                \sffamily\fontsize{60}{60}\selectfont%
                \color{white}%
                \thechapter%
            };%
            \node [inner sep = 0, outer sep = 0, anchor = north west] at (numbernode.south west)%
            {%
                \begin{tikzpicture}[overlay, remember picture]%
                    \draw[color = stroke1, line width = 0.7 mm] (\myinnermargin, -1 ex) -- (\paperwidth, -1 ex);%
                \end{tikzpicture}%
            };%
            \node (align) [align = left, right = \myinnermargin, inner sep = 0, outer sep = 0, anchor = south west] at (numbernode.south west)%
            {%
                #3%
            };%
        \end{tikzpicture}%
    }%
}
\RedeclareSectionCommand%
[%
    font = \color{stroke1}\normalfont\huge\sffamily,
    afterskip = 20 pt,
]
{chapter}

\BeforeStartingTOC[toc]{\pagestyle{plain}}
\AfterStartingTOC{\thispagestyle{plain}}

%% file: core/bibliography/bibliography_format.tex
\usepackage
[
    sortcites,              
    style = alphabetic,     
    defernumbers,           
    safeinputenc,           
    backref = true,         
    backrefstyle = three,   
    hyperref = true,        
    maxbibnames = 99,       
    maxcitenames = 2,       
]
{biblatex} 

\renewbibmacro{in:}%
{%
    \ifentrytype{article}{}{\printtext{\bibstring{in}\intitlepunct}}%
}

\renewbibmacro*{volume+number+eid}%
{%
    \printfield{volume}%
    \iffieldundef{number}{}{\addcolon}%
    \printfield{number}%
    \setunit*{\addcomma\space}%
    \printfield{eid}%
}

\DefineBibliographyStrings{english}%
{%
    backrefpage  = {\lowercase{s}ee page}, 
    backrefpages = {\lowercase{s}ee pages} 
}

\DeclareFieldFormat[article]{title}{\textbf{\color{stroke1}#1}}
\DeclareFieldFormat[inproceedings]{title}{\textbf{\color{stroke1}#1}}
\DeclareFieldFormat[thesis]{title}{\textbf{\color{stroke1}#1}}
\DeclareFieldFormat[book]{title}{\textbf{\color{stroke1}#1}}
\DeclareFieldFormat[unpublished]{title}{\textbf{\color{stroke1}#1}}
\DeclareFieldFormat[report]{title}{\textbf{\color{stroke1}#1}}
\DeclareFieldFormat[inbook]{chapter}{\textbf{\color{stroke1}#1}}
\DeclareFieldFormat[inbook]{title}{#1}
\DeclareFieldFormat{pages}{#1}

\newtoggle{authorend}
\togglefalse{authorend}

\DeclareBibliographyDriver{article}%
{%
  \usebibmacro{bibindex}%
  \usebibmacro{begentry}%
  \iftoggle{authorend}{}{\usebibmacro{author/translator+others}}%
  \setunit{\labelnamepunct}\newblock
  \usebibmacro{title}%
  \newunit
  \printlist{language}%
  \newunit\newblock
  \usebibmacro{byauthor}%
  \newunit\newblock
  \usebibmacro{bytranslator+others}%
  \newunit\newblock
  \printfield{version}%
  \newunit\newblock
  \usebibmacro{in:}%
  \usebibmacro{journal+issuetitle}%
  \newunit
  \usebibmacro{byeditor+others}%
  \newunit
  \usebibmacro{note+pages}%
  \newunit\newblock
  \iftoggle{bbx:isbn}
  {\printfield{issn}}
  {}%
  \newunit\newblock
  \usebibmacro{doi+eprint+url}%
  \newunit\newblock
  \usebibmacro{addendum+pubstate}%
  \setunit{\bibpagerefpunct}\newblock
  \usebibmacro{pageref}%
  \newunit\newblock
  \iftoggle{bbx:related}
  {\usebibmacro{related:init}%
    \usebibmacro{related}}
  {}%
  \usebibmacro{finentry}%
  \iftoggle{authorend}{\usebibmacro{author/translator+others}}{}%
}

\DeclareBibliographyDriver{inbook}%
{%
  \usebibmacro{bibindex}%
  \usebibmacro{begentry}%
  \iftoggle{authorend}{}{\usebibmacro{author/translator+others}}%
  \setunit{\labelnamepunct}\newblock
  \usebibmacro{chapter+pages}%
  \newunit
  \printlist{language}%
  \newunit\newblock
  \usebibmacro{byauthor}%
  \newunit\newblock
  \usebibmacro{in:}%
  \usebibmacro{bybookauthor}%
  \newunit\newblock
  \usebibmacro{maintitle+booktitle}%
  \newunit\newblock
  \usebibmacro{byeditor+others}%
  \newunit\newblock
  \printfield{edition}%
  \newunit
  \iffieldundef{maintitle}
  {\printfield{volume}%
    \printfield{part}}
  {}%
  \newunit
  \printfield{volumes}%
  \newunit\newblock
  \usebibmacro{series+number}%
  \newunit\newblock
  \printfield{note}%
  \newunit\newblock
  \usebibmacro{publisher+location+date}%
  \newunit\newblock
  \newunit\newblock
  \iftoggle{bbx:isbn}
  {\printfield{isbn}}
  {}%
  \newunit\newblock
  \usebibmacro{doi+eprint+url}%
  \newunit\newblock
  \usebibmacro{addendum+pubstate}%
  \setunit{\bibpagerefpunct}\newblock
  \usebibmacro{pageref}%
  \newunit\newblock
  \iftoggle{bbx:related}
  {\usebibmacro{related:init}%
    \usebibmacro{related}}
  {}%
  \usebibmacro{finentry}%
  \iftoggle{authorend}{\usebibmacro{author/translator+others}}{}%
}

\DeclareBibliographyDriver{inproceedings}%
{%
  \usebibmacro{bibindex}%
  \usebibmacro{begentry}%
  \iftoggle{authorend}{}{\usebibmacro{author/translator+others}}%
  \setunit{\labelnamepunct}\newblock
  \usebibmacro{title}%
  \newunit
  \printlist{language}%
  \newunit\newblock
  \usebibmacro{byauthor}%
  \newunit\newblock
  \usebibmacro{in:}%
  \usebibmacro{maintitle+booktitle}%
  \newunit\newblock
  \usebibmacro{event+venue+date}%
  \newunit\newblock
  \usebibmacro{byeditor+others}%
  \newunit\newblock
  \iffieldundef{maintitle}
  {\printfield{volume}%
    \printfield{part}}
  {}%
  \newunit
  \printfield{volumes}%
  \newunit\newblock
  \usebibmacro{series+number}%
  \newunit\newblock
  \printfield{note}%
  \newunit\newblock
  \printlist{organization}%
  \newunit
  \usebibmacro{publisher+location+date}%
  \newunit\newblock
  \usebibmacro{chapter+pages}%
  \newunit\newblock
  \iftoggle{bbx:isbn}
  {\printfield{isbn}}
  {}%
  \newunit\newblock
  \usebibmacro{doi+eprint+url}%
  \newunit\newblock
  \usebibmacro{addendum+pubstate}%
  \setunit{\bibpagerefpunct}\newblock
  \usebibmacro{pageref}%
  \newunit\newblock
  \iftoggle{bbx:related}
  {\usebibmacro{related:init}%
    \usebibmacro{related}}
  {}%
  \usebibmacro{finentry}%
  \iftoggle{authorend}{\usebibmacro{author/translator+others}}{}%
}

\DeclareBibliographyDriver{thesis}%
{%
  \usebibmacro{bibindex}%
  \usebibmacro{begentry}%
  \iftoggle{authorend}{}{\usebibmacro{author}}%
  \setunit{\labelnamepunct}\newblock
  \usebibmacro{title}%
  \newunit
  \printlist{language}%
  \newunit\newblock
  \usebibmacro{byauthor}%
  \newunit\newblock
  \printfield{note}%
  \newunit\newblock
  \printfield{type}%
  \newunit
  \usebibmacro{institution+location+date}%
  \newunit\newblock
  \usebibmacro{chapter+pages}%
  \newunit
  \printfield{pagetotal}%
  \newunit\newblock
  \iftoggle{bbx:isbn}
  {\printfield{isbn}}
  {}%
  \newunit\newblock
  \usebibmacro{doi+eprint+url}%
  \newunit\newblock
  \usebibmacro{addendum+pubstate}%
  \setunit{\bibpagerefpunct}\newblock
  \usebibmacro{pageref}%
  \newunit\newblock
  \iftoggle{bbx:related}
  {\usebibmacro{related:init}%
    \usebibmacro{related}}
  {}%
  \usebibmacro{finentry}%
  \iftoggle{authorend}{\usebibmacro{author}}{}%
}

\providebool{bbx:subentry}
\newbibmacro*{citenum}%
{
  \printtext[bibhyperref]{
    \printfield{prefixnumber}%
    \printfield{labelnumber}%
    \ifbool{bbx:subentry}
    {\printfield{entrysetcount}}
    {}}%
}

\DeclareCiteCommand{\conline}[\mkbibbrackets]
{\usebibmacro{prenote}}
{\usebibmacro{citeindex}%
  \usebibmacro{citenum}}
{\multicitedelim}
{\usebibmacro{postnote}}

%% file: packages_and_commands/standard_packages.tex
\usepackage
[
    babel = true, 
]
{microtype}           
\usepackage{csquotes} 

\usepackage{amsmath}
\usepackage{amssymb}
\usepackage{amsthm}
\usepackage{thmtools}
\usepackage{mathtools}
\usepackage{thm-restate}
\usepackage{dsfont}        
\usepackage{braceMnSymbol} 

\usepackage
[
    ttscale = 0.85, 
]
{libertine} 
\usepackage
[
    libertine,    
    slantedGreek, 
    vvarbb,       
    libaltvw,     
]
{newtxmath} 
\usepackage{url} 
\usepackage{bm}  

\usepackage{graphicx} 
\usepackage
[
    subrefformat = simple, 
    labelformat = simple,  
]
{subcaption}         
\usepackage{wrapfig} 

\columnsep = \mymargininnersep

\usepackage{array}     
\usepackage{booktabs}  
\usepackage{longtable} 
\usepackage{pdflscape} 

\usepackage{enumitem} 

\usepackage
[
    ruled,         
    vlined,        
    linesnumbered, 
]
{algorithm2e} 

\usepackage{xspace}   
\usepackage
[
    shortcuts, 
]
{extdash}             
\usepackage{setspace} 

\usepackage{xparse}    
\usepackage{footnote}  
\usepackage{afterpage} 
\usepackage
[
    textsize = scriptsize, 
]
{todonotes}            


\usepackage
[
    bookmarks = true,                 
    bookmarksopen = false,            
    bookmarksnumbered = true,         
    pdfstartpage = 1,                 
    pdftitle = {{\printTitle}},       
    pdfauthor = {{\printAuthor}},     
    pdfsubject = {{\printSubject}},   
    pdfkeywords = {{\printKeywords}}, 
    breaklinks = true,                
    \ifprintVersion
        hidelinks,                    
    \else
    colorlinks = true,            
    allcolors = stroke1,          
    \fi
]
{hyperref} 

\usepackage
[
    noabbrev,   
    nameinlink, 
]
{cleveref} 

%% file: packages_and_commands/general_commands.tex
\newcommand*{\colloquialDegreeName}{Master}
\newcommand*{\colloquialDegreeNameLowercase}{master}

\newcommand*{\degreeAbbreviation}{M.}

\ifbachelorThesis
    \renewcommand*{\colloquialDegreeName}{Bachelor}
    \renewcommand*{\colloquialDegreeNameLowercase}{bachelor}
    \renewcommand*{\degreeAbbreviation}{B.}
\fi

\makeatletter
    \def\IfEmptyTF#1%
    {%
        \if\relax\detokenize{#1}\relax%
            \expandafter\@firstoftwo%
        \else%
            \expandafter\@secondoftwo%
        \fi%
    }
\makeatother

\NewDocumentCommand{\mathOrText}{m}
{%
    \ensuremath{#1}\xspace%
}

\let\originalleft\left
\let\originalright\right
\renewcommand{\left}{\mathopen{}\mathclose\bgroup\originalleft}
\renewcommand{\right}{\aftergroup\egroup\originalright}

\makeatletter
    \DeclareRobustCommand{\bfseries}%
    {%
        \not@math@alphabet\bfseries\mathbf%
        \fontseries\bfdefault\selectfont%
        \boldmath%
    }
\makeatother

\xspaceaddexceptions{]\}}

\allowdisplaybreaks

\crefname{ineq}{inequality}{inequalities}
\creflabelformat{ineq}{#2{\upshape(#1)}#3} 

\crefname{term}{term}{terms}
\creflabelformat{term}{#2{\upshape(#1)}#3}

\let\oldfootnote\footnote

\newlength{\spaceBeforeFootnote} 
\newlength{\spaceAfterFootnote}  

\RenewDocumentCommand{\footnote}{o o o m}%
{%
    \IfNoValueTF{#1}%
    {%
        \oldfootnote{#4}%
    }%
    {%
        \setlength{\spaceBeforeFootnote}{\IfEmptyTF{#1}{0}{#1} em}%
        \IfNoValueTF{#2}%
        {%
            \hspace*{\spaceBeforeFootnote}\oldfootnote{#4}%
        }%
        {%
            \setlength{\spaceAfterFootnote}{\IfEmptyTF{#2}{0}{#2} em}%
            \hspace*{\spaceBeforeFootnote}\IfNoValueTF{#3}{\oldfootnote{#4}}{\oldfootnote[#3]{#4}}\hspace*{\spaceAfterFootnote}%
        }%
    }%
}

\makesavenoteenv{figure}
\makesavenoteenv{table}
\makesavenoteenv{tabular}

\iffancyTheorems
    \declaretheoremstyle
    [
        spaceabove = \topsep,
        spacebelow = \topsep,
        headfont = \bfseries,
        headformat = \textcolor{stroke1}{$\blacktriangleright$} \NAME~\NUMBER \NOTE,
        notefont = \bfseries,
        notebraces = {(}{)},
        bodyfont = \normalfont,
        postheadspace = 0.5 em,
        qed = \textcolor{stroke1}{\bfseries$\blacktriangleleft$},
    ]
    {myTheoremStyle}
    
    \declaretheorem
    [
        style = myTheoremStyle,
        name = Conjecture,
        numberwithin = chapter,
    ]
    {conjecture}
    \declaretheorem
    [
        style = myTheoremStyle,
        name = Proposition,
        sharenumber = conjecture,
    ]
    {proposition}
    \declaretheorem
    [
        style = myTheoremStyle,
        name = Claim,
        sharenumber = conjecture,
    ]
    {claim}
    \declaretheorem
    [
        style = myTheoremStyle,
        name = Lemma,
        sharenumber = conjecture,
    ]
    {lemma}
    \declaretheorem
    [
        style = myTheoremStyle,
        name = Corollary,
        sharenumber = conjecture,
    ]
    {corollary}
    \declaretheorem
    [
        style = myTheoremStyle,
        name = Theorem,
        sharenumber = conjecture,
    ]
    {theorem}
    \declaretheorem
    [
        style = myTheoremStyle,
        name = Definition,
        sharenumber = conjecture,
    ]
    {definition}
    \declaretheorem
    [
        style = myTheoremStyle,
        name = Example,
        sharenumber = conjecture,
    ]
    {example}
    \declaretheorem
    [
        style = myTheoremStyle,
        name = Remark,
        sharenumber = conjecture,
    ]
    {remark}
\else
    \theoremstyle{plain}
    
    \newtheorem{conjecture}{Conjecture}[chapter]

    \newtheorem{lemma}[conjecture]{Lemma}
    \newtheorem{corollary}[conjecture]{Corollary}
    \newtheorem{theorem}[conjecture]{Theorem}
    \newtheorem{definition}[conjecture]{Definition}

\fi

\NewDocumentCommand{\functionTemplate}{m m m m o}%
{%
    \IfNoValueTF{#5}%
    {%
        \mathOrText{#1\left#2{#4}\right#3}%
    }%
    {%
        \mathOrText{#1#5#2{#4}#5#3}%
    }%
}

\newcommand*{\leftBracketType}{(}
\newcommand*{\rightBracketType}{)}

\NewDocumentCommand{\createFunction}{m m o o}%
{%
    \renewcommand*{\leftBracketType}{\IfNoValueTF{#3}{(}{#3}}%
    \renewcommand*{\rightBracketType}{\IfNoValueTF{#4}{)}{#4}}%
    \NewDocumentCommand{#1}{o o}%
    {%
        \IfNoValueTF{##1}%
        {%
            \mathOrText{#2}%
        }%
        {%
            \functionTemplate{#2}{\leftBracketType}{\rightBracketType}{##1}[##2]%
        }%
    }%
}

\DeclareDocumentCommand{\probabilisticFunctionTemplate}{m m O{} o}
{%
    \functionTemplate{#1}%
    {\lbrack}%
    {\rbrack}%
    {#2\IfEmptyTF{#3}{}{\ \IfNoValueTF{#4}{\left}{#4}\vert\ \vphantom{#2}#3\IfNoValueTF{#4}{\right.}{}}}%
    [#4]%
}

\ifboldNumberSets
    \newcommand*{\N}{\mathOrText{\mathbf{N}}}

    \newcommand*{\indicatorFunctionSymbol}{\mathbf{1}}
\else
    \newcommand*{\N}{\mathOrText{\mathds{N}}}

    \newcommand*{\indicatorFunctionSymbol}{\mathds{1}}
\fi

\RenewDocumentCommand{\Pr}{m O{} o}%
{%
    \probabilisticFunctionTemplate{\mathrm{Pr}}{#1}[#2][#3]%
}

\NewDocumentCommand{\E}{m O{} o}%
{%
    \probabilisticFunctionTemplate{\mathrm{E}}{#1}[#2][#3]%
}

\NewDocumentCommand{\Var}{m O{} o}%
{%
    \probabilisticFunctionTemplate{\mathrm{Var}}{#1}[#2][#3]%
}

\DeclareDocumentCommand{\bigO}{m o}%
{%
    \functionTemplate{\mathrm{O}}{(}{)}{#1}[#2]%
}

\DeclareDocumentCommand{\smallO}{m o}%
{%
    \functionTemplate{\mathrm{o}}{(}{)}{#1}[#2]%
}

\DeclareDocumentCommand{\bigTheta}{m o}%
{%
    \functionTemplate{\upTheta}{(}{)}{#1}[#2]%
}

\DeclareDocumentCommand{\bigOmega}{m o}%
{%
    \functionTemplate{\upOmega}{(}{)}{#1}[#2]%
}

\DeclareDocumentCommand{\smallOmega}{m o}%
{%
    \functionTemplate{\upomega}{(}{)}{#1}[#2]%
}

\DeclareDocumentCommand{\eulerE}{o}%
{%
    \mathOrText{\mathrm{e}\IfNoValueTF{#1}{}{^{#1}}}%
}

\DeclareDocumentCommand{\poly}{m o}%
{%
    \functionTemplate{\mathrm{poly}}{(}{)}{#1}[#2]%
}

\createFunction{\id}{\mathrm{id}}

\NewDocumentCommand{\indic}{m o o}%
{%
    \IfNoValueTF{#2}%
    {%
        \mathOrText{\indicatorFunctionSymbol_{#1}}%
    }%
    {%
        \functionTemplate{\indicatorFunctionSymbol_{#1}}{(}{)}{#2}[#3]%
    }%
}

\DeclareDocumentCommand{\rng}{m o}%
{%
    \functionTemplate{\mathrm{rng}}{(}{)}{#1}[#2]%
}

\DeclareDocumentCommand{\d}{o}%
{%
    \mathrm{d}\IfNoValueTF{#1}{}{^{#1}}%
}


%% file: packages_and_commands/user-defined_commands.tex
\DeclareMathSymbol{:}{\mathpunct}{operators}{"3A}

\newcommand{\cardinality}[1]{\vert #1 \vert}

\newcommand{\with}{\text{ } | \text{ }}

\newcommand{\Iff}{\Leftrightarrow}
\newcommand{\Implies}{\Rightarrow}

\usepackage{tikz}
\usetikzlibrary{shapes,fit, decorations.pathreplacing}

\newcommand*{\Txt}{\mathbf{Txt}}
\newcommand*{\Inf}{\mathbf{Inf}}
\newcommand*{\InfCan}{\Inf_{\textup{can}}}

\newcommand*{\G}{\mathbf{G}}
\newcommand*{\CIt}{\mathbf{CflIt}}
\newcommand*{\It}{\mathbf{It}}
\newcommand*{\Sd}{\mathbf{Sd}}
\newcommand*{\Psd}{\mathbf{Psd}}

\newcommand*{\Ex}{\mathbf{Ex}}
\newcommand*{\Bc}{\mathbf{Bc}}

\newcommand*{\Cons}{\mathbf{Cons}}

\newcommand*{\Caut}{\mathbf{Caut}}
\newcommand*{\CautTar}{\Caut_{\textup{\textbf{Tar}}}}
\newcommand*{\CautFin}{\Caut_{\textup{\textbf{Fin}}}}
\newcommand*{\CautInf}{\Caut_{\infty}}
\newcommand*{\True}{\mathbf{T}}
\newcommand*{\Wb}{\mathbf{Wb}}
\newcommand*{\NU}{\mathbf{NU}}

\newcommand*{\Dec}{\mathbf{Dec}}

\newcommand*{\WMon}{\mathbf{WMon}}
\newcommand*{\Mon}{\mathbf{Mon}}
\newcommand*{\SMon}{\mathbf{SMon}}
\newcommand*{\DWMon}{\mathbf{WMon}^d}
\newcommand*{\DMon}{\mathbf{Mon}^d}
\newcommand*{\DSMon}{\mathbf{SMon}^d}
\newcommand*{\BWMon}{\mathbf{WMon}^{\&}}
\newcommand*{\BMon}{\mathbf{Mon}^{\&}}
\newcommand*{\BSMon}{\mathbf{SMon}^{\&}}

\newcommand*{\Sem}{\mathbf{Sem}}

\newcommand*{\SemWb}{\Sem\Wb}

\newcommand{\set}[1]{\{ #1 \}}
\newcommand*{\bset}{\set{0, 1}{}}
\newcommand*{\ncmp}{\#}

\newcommand*{\La}{\mathcal{L}}

\newcommand*{\Seq}{{\mathbb{S}\mathrm{eq}}}

\newcommand*{\totalFn}{\mathfrak{R}}
\newcommand*{\partialFn}{\mathfrak{P}}
\newcommand*{\totalCp}{\mathcal{R}}
\newcommand*{\partialCp}{\mathcal{P}}
\newcommand*{\enum}{\mathcal{E}}

\newcommand*{\simulatingFn}{\mathfrak{S}}

\newcommand{\dom}{\mathrm{dom}}
\newcommand{\range}{\mathrm{range}}
\newcommand{\content}{\mathrm{content}}
\newcommand{\outline}{\mathrm{outline}}

\newcommand{\pos}{\mathrm{pos}}
\renewcommand{\neg}{\mathrm{neg}} 

\newcommand{\patch}{\mathrm{patch}}
\newcommand{\suc}{\mathrm{succ}}

\newcommand*{\subSeq}{\sqsubseteq}
\def\sqsubsetneq{\mathrel{\sqsubseteq\kern-0.735em\raise-0.07em\hbox{\rotatebox{313}{\scalebox{0.7}[0.5]{\(\shortmid\)}}}\scalebox{0.35}[1]{\ }}}
\newcommand*{\subSeqNeq}{\sqsubsetneq}

\newcommand*{\trans}{\preccurlyeq}

\newcommand*{\semTrans}{\preccurlyeq_{\textup{sem}}}

\newcommand*{\semTransEq}{\cong_{\textup{sem}}}

\newcommand{\semequiv}{\equiv_W}

\newcommand*{\ORT}{\textbf{ORT}\xspace}
\newcommand*{\KRT}{\textbf{KRT}\xspace}

\newcommand*{\convs}{{\downarrow}}
\newcommand*{\divs}{{\uparrow}}

\makeatletter
\newsavebox{\@brx}
\newcommand{\llangle}[1][]{\savebox{\@brx}{\(\m@th{#1\langle}\)}%
  \mathopen{\copy\@brx\kern-0.5\wd\@brx\usebox{\@brx}}}
\newcommand{\rrangle}[1][]{\savebox{\@brx}{\(\m@th{#1\rangle}\)}%
  \mathclose{\copy\@brx\kern-0.5\wd\@brx\usebox{\@brx}}}
\makeatother

\pgfdeclarelayer{background}
\pgfdeclarelayer{foreground}
\pgfsetlayers{background,main,foreground}

%% file: core/title_page/title_page.tex

\ifprintVersion
    \ifprofessionalPrint
        \newgeometry
        {
            textwidth = 134 mm,
            textheight = 220 mm,
            top = 38 mm + \extraborderlength,
            inner = 38 mm + \mybindingcorrection + \extraborderlength,
        }
    \else
        \newgeometry
        {
            textwidth = 134 mm,
            textheight = 220 mm,
            top = 38 mm,
            inner = 38 mm + \mybindingcorrection,
        }
    \fi
\else
    \newgeometry
    {
        textwidth = 134 mm,
        textheight = 220 mm,
        top = 38 mm,
        inner = 38 mm,
    }
\fi

\begin{titlepage}
    \sffamily
    \begin{center}
        \includegraphics[height = 3.2 cm]{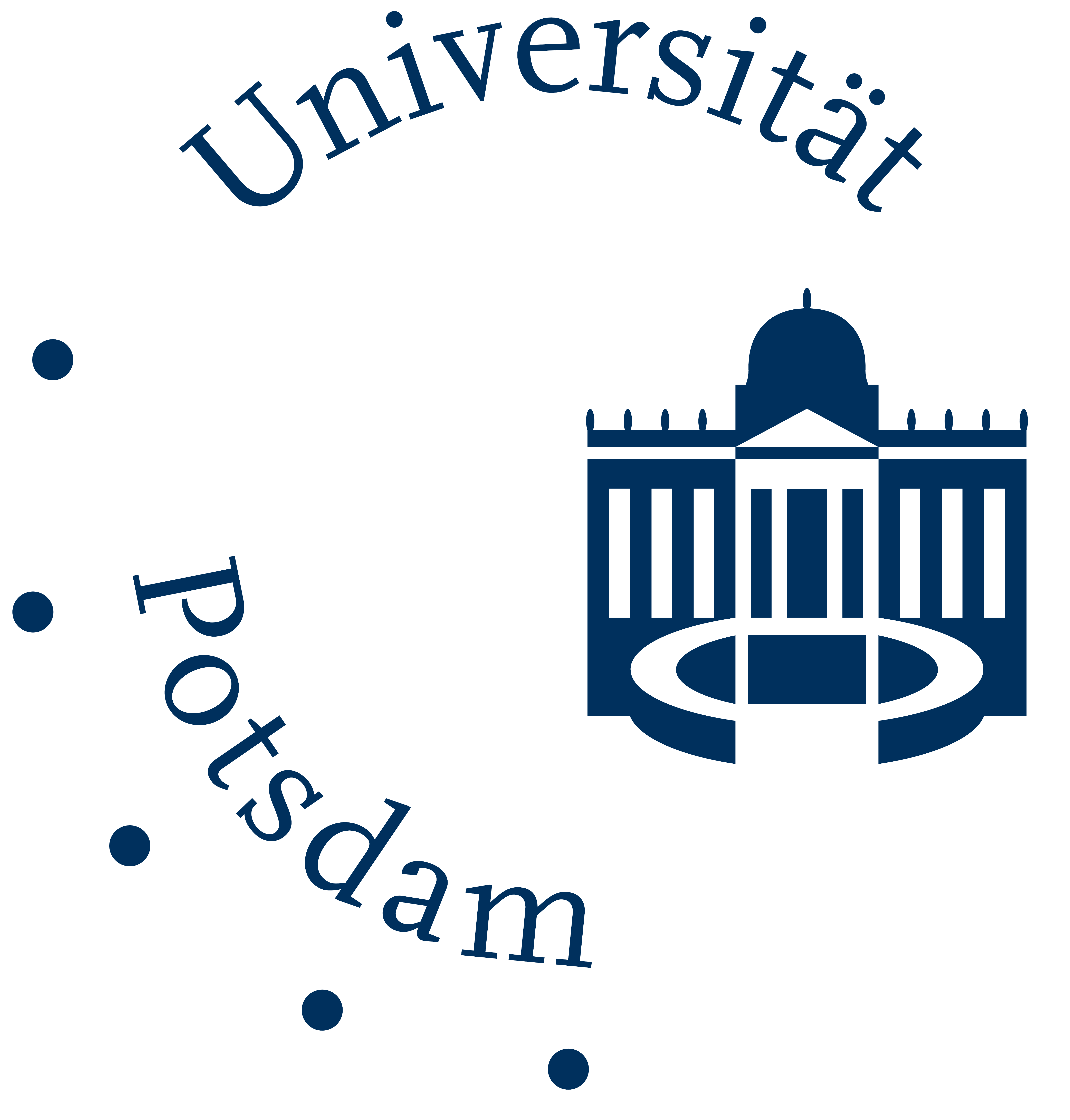} \hfill \includegraphics[height = 3 cm]{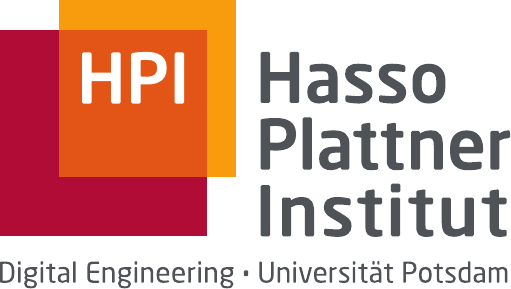}\\
        \vfil
        {\LARGE
            \rule[1 ex]{\textwidth}{1.5 pt}
            \onehalfspacing\printTitleBold\\[1 ex]
            {\vspace*{-1 ex}\Large \printGermanTitle}\\
            \rule[-1 ex]{\textwidth}{1.5 pt}
        }
        \vfil
        {\Large\textbf{\printAuthor}}
        \vfil
        {\large Universitäts\colloquialDegreeNameLowercase arbeit\\[0.25 ex]
        zur Erlangung des akademischen Grades}\\[0.25 ex]
        \bigskip
        {\Large \colloquialDegreeName{} of Science}\\[0.5 ex]
        {\large\emph{(\degreeAbbreviation\,Sc.)}}\\
        \bigskip
        {\large im Studiengang\\[0.25 ex]
        \printProgram}
        \vfil
        {\large eingereicht am \printDateReceived{} am\\[0.25 ex]
        Fachgebiet Algorithm Engineering der\\[0.25 ex]
        Digital-Engineering-Fakultät\\[0.25 ex]
        der Universität Potsdam}
    \end{center}
    
    \vfil
    \begin{table}[h]
        \centering
        \large
        \sffamily 
        {\def\arraystretch{1.2}
            \begin{tabular}{>{\bfseries}p{3.8 cm}p{5.3 cm}}
                Gutachter               & \printNameOfSupervisor\\
                Betreuer                & \printAdditionalExaminers
            \end{tabular}
        }
    \end{table}
\end{titlepage}

\restoregeometry

%% file: preface/abstract.tex
In inductive inference, we investigate the learnability of classes of formal languages.
We are interested in what classes of languages are learnable in certain learning settings.
A class of languages is learnable, if there is a learner that can identify all of its languages and satisfies the constraints of the learning setting.
To identify a language, a learner is presented with information about this very language.
When learning from informants, this information consists of examples for numbers that are, and numbers that are not included in the target language.
As more and more examples are presented, the learner outputs a hypothesis sequence.
To satisfy behaviorally correct identification, this hypothesis sequence must eventually only list correct labels for the target language.
In this thesis, we compare the effects of a number of semantic learning restrictions on the learning capabilities for behaviorally correct learning from informants.

To start, we collect and combine some known theorems to show that we can assume learners to be set-driven and total.
Additionally, learners only need to identify languages by their canonical informant, an informant that is particularly well-formed.
We then investigate the effects of a number of learning restrictions on the inference capabilities of learners.
Most importantly, we specify all relations between monotonic and strong monotonic restrictions, including dual and combined versions.
Monotonicity restrictions require that a learner outputs better generalizations (or, for the dual variants, specializations) over time.
Similarly to what has been found in the setting of learning indexed families, these restrictions form a strict hierarchy.
We show that weak monotonicity does not restrict learners.
We reprove that cautiousness is a proper restriction by investigating three weaker variants.
Interestingly, while infinite cautiousness does not lessen learning power in full-information text-learning, the setting they were introduced in, we show that all three decrease the set of learnable classes in our setting.
Finally, we show that we can assume global consistency for learners satisfying any of the monotonic restrictions we investigate, with the exception of combined weak monotonicity.

%% file: preface/abstractGerman.tex
In der induktiven Inferenz untersuchen wir die Lernbarkeit von Klassen formaler Sprachen.
Wir sind daran interessiert, welche Klassen von Sprachen in bestimmten Lernumgebungen lernbar sind.
Eine Klasse von Sprachen ist lernbar, wenn es einen Lerner gibt, der alle enthaltenen Sprachen identifizieren kann und die Bedingungen der Lernumgebung erfüllt.
Um eine Sprache zu identifizieren, erhält der Lerner Informationen über ebendiese Sprache.
Beim Lernen von Informanten bestehen diese Informationen aus Beispielen für Zahlen, die in der Zielsprache enthalten sind, und für Zahlen, die in der Zielsprache nicht enthalten sind.
Während mehr und mehr Beispiele präsentiert werden, gibt der Lerner eine Hypothesenfolge aus.
Für verhaltenskorrekte Identifikation muss diese Hypothesenfolge irgendwann nur noch die Zielsprache beschreiben.
In dieser Arbeit vergleichen wir die Auswirkungen einer Reihe semantischer Lernrestriktionen auf die Lernfähigkeit für verhaltenskorrektes Lernen von Informanten.

Zu Beginn sammeln und kombinieren wir einige bekannte Theoreme, um zu zeigen, dass wir davon ausgehen können, dass Lerner mengengetrieben und total sind.
Außerdem müssen Lerner Sprachen nur anhand ihres kanonischen Informanten identifizieren, also eines Informanten, der besonders wohlgeformt ist.
Anschließend untersuchen wir die Auswirkungen einer Reihe von Lernrestriktionen auf die Inferenzfähigkeiten des Lerners.
Vor allem spezifizieren wir alle Beziehungen zwischen monotonen und stark monotonen Restriktionen, einschließlich dualer und kombinierter Varianten.
Monotonitätsrestriktionen setzen voraus, dass ein Lerner im Laufe der Zeit bessere Verallgemeinerungen (oder, bei den dualen Varianten, Spezialisierungen) liefert.
Ähnlich wie beim Lernen indizierter Familien bilden diese Einschränkungen eine strenge Hierarchie.
Wir zeigen, dass schwache Monotonität Lerner nicht einschränkt.
Wir beweisen, dass Behutsamkeit eine tatsächliche Einschränkung ist, indem wir drei schwächere Varianten untersuchen.
Obwohl unendliche Behutsamkeit die Lernfähigkeit beim Text-Lernen mit vollständiger Information nicht verringert, zeigen wir, dass alle drei die Menge der lernbaren Klassen in unserer Konfiguration verringern.
Schließlich zeigen wir, dass wir globale Konsistenz für Lerner annehmen können, die alle von uns untersuchten monotonen Restriktionen erfüllen, mit Ausnahme der kombinierten schwachen Monotonität.

%% file: introduction/introduction.tex
\emph{Inductive inference}, also called \emph{language learning in the limit}, is a branch of \emph{recursion theory} and was introduced by \textcite{gold1967language}.
The model was initially designed to draw parallels to how humans learn natural languages.
Nowadays we are more interested into its implications for computability theory, machine learning and binary classification \cite{seidel2021modelling}.
Inductive inference is about the learnability of classes of recursively enumerable \emph{formal languages}, i.e.\ subsets of the natural numbers.
A class of languages is \emph{learnable}, if there is an algorithmic \emph{learner} that can correctly identify all of its languages.
To identify a language, the learner has to recognize it when given some hints about it.
We study what classes of languages are learnable when varying our requirements for admissible learners.
For example, the content of the given information depends on the \emph{presentation system}.
The two presentation systems for language learning are \emph{text} ($\Txt$) and \emph{informant} ($\Inf$), both of which have been introduced by \textcite{gold1967language}.
A text contains elements of the target language, whereas an informant also gives counter-examples.

\emph{Convergence criteria} define what it means to correctly identify a language.
The default convergence criterion is \emph{explanatory learning} ($\Ex$) and was introduced by \textcite{gold1967language}.
$\Ex$ requires that the learner converges to one output that correctly explains the target language.
A relaxed version of $\Ex$ is \emph{behaviorally correct} ($\Bc$) identification \cite{case1983comparison}.
$\Bc$-learners are allowed to output different explanations as long as they all explain the correct language.
Out of the four combinations of text- and informant-learning and explanatory and behaviorally correct identification, $\Bc$-learning from informants has been studied the least.
In this thesis we collect known theorems in this area and fill some of the gaps.

Fixing $\Inf$ and $\Bc$, we compare the effect of \emph{learning restrictions} on the inference capabilities of learners.
As they are given more and more information, learners output a \emph{hypothesis sequence}.
Learning restrictions are predicates over the hypothesis sequence and give a formal model for requiring learners to, for example, come up with better generalizations over time, come up with better specializations over time or just not contradict the input data.
Learning restrictions can be grouped by some properties.
A learning restriction is called \emph{semantic}, if it is only concerned about a learner's conjectured languages, not the labels used to encode those \cite{kotzing2017solution}.
Another property of learning restrictions is \emph{delayability}, which requires that the restriction is preserved when shifting or skipping hypotheses \cite{kotzing2016map}.

\section{Contributions}

We arrange our results in four chapters: The first for general observations and the other three for cautiousness, monotonicity and consistency.

The general observations are collected and combined theorems for groups of restrictions.
\begin{itemize}[noitemsep,nosep]
    \item For delayable restrictions, it is sufficient that learners identify languages by their canonical (a particularly well-formed) informant and that we can assume them to be set-driven and total (see \Cref{thm:totalInteractionForDelayable}).
    \item For restrictions that are both delayable and semantic, we can additionally limit the interaction with a learner to be iterative (see \Cref{thm:interactionForDelaySem}).
\end{itemize}
Using these findings, we show that the relaxation of $\Bc$ from $\Ex$ actually increases inference capabilities of learners (see \Cref{thm:bcexsep}).

We reprove that cautiousness is a proper restriction by investigating three weaker variants.
We find that all three properly restrict learners in our setting (see \Cref{thm:cauttarandinfproper}).
This is in contrast to explanatory learning form text, where infinite cautiousness does not weaken learners \cite{kotzing2016map}.

\begin{figure}[h]
    \centering
    \input{introduction/map}
    \caption{
        Relations between monotonicity constraints in $\Inf\Bc$-learning.
        Black lines indicate inclusions.
        Two learning restrictions are equivalent if and only if they lie in the same gray box.
        For all displayed restrictions, learners can additionally be assumed to be set-driven, total and globally consistent.
    }
    \label{fig:monmap}
\end{figure}
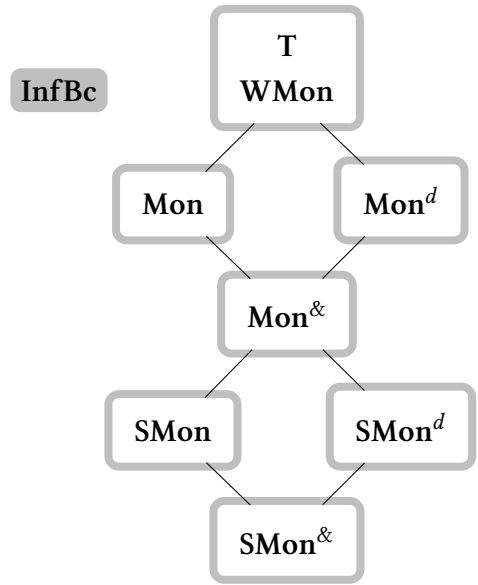

We provide an initial map for behaviorally correct learning from informants that covers monotonic and strong monotonic restrictions and their dual and combined counterparts (see \Cref{fig:monmap}).
We find that these restrictions form a strict hierarchy, analogous to what has been observed in the adjacent setting of learning \emph{indexed families} by \textcite{lange1994characterization}.
This is due to the fact that most of their separations can be transferred while the inclusions hold by definition.
We observe that weak monotonicity, like with $\Ex$, does not pose a proper restriction on learners.
However, our approach is very different from constructions for similar theorems in explanatory learning or learning indexed families.
Additionally, our constructed learner behaves globally consistent (see \Cref{thm:conswmonistrue}).

In \Cref{thm:InfBcConsAllEq}, \Cref{thm:patchPreservesBDSMon} and \Cref{thm:makedualwmoncons}, we construct globally consistent learners that preserve
\begin{itemize}[noitemsep,nosep]
    \item behaviorally correct identification,
    \item classic, dual and combined monotonicity,
    \item classic, dual and combined strong monotonicity and
    \item dual weak monotonicity.
\end{itemize}

Throughout the thesis, we employ established proof techniques such as separation using the \emph{Operator Recursion Theorem} \cite{case1994infinitary} (see \Cref{thm:bcexsep}) and \emph{poisoning} (see \Cref{thm:conswmonistrue}).

\section{Related work}

The most exhaustive work on learning from informants is by \textcite{aschenbach2018learning}, where a map for a lot of common learning restrictions in the $\Ex$ setting is provided.
Although they did not explicitly consider $\Bc$, some of their separations transfer directly into our setting.
Interestingly, they found that only $\Mon$, $\Caut$ and $\SMon$ impose a proper restriction on learners.
Furthermore, they observed that learners may be assumed to be set-driven.
This is in contrast to text learning, where set-drivenness is a strong restriction.
Their map does not include dual monotonic learning or the weak and strong counterparts.
For learning indexed families, \textcite{lange1996monotonic} built a map for all monotonic restrictions in an $\Inf\Ex$ learning scenario.
Monotonic learning has also been studied in other research, for example by \textcite{doskovc2021mapping}.

Concerning behaviorally correct learning, most findings are included in more general theorems about delayable or semantic restrictions.
As an example, \textcite{aschenbach2018learning} showed that totality is not a restrictive assumption for delayable learning restrictions.
Totality can be assumed for all semantic restrictions too, as shown by \textcite{kotzing2017normal}.

Since most of the learning restrictions were built with $\Ex$-learning in mind, there has been work to develop equivalent restrictions that are semantic.
\Textcite{kotzing2017normal} introduced the concept of a \emph{semantic closure} to better classify restrictions and to find semantic equivalents of known restrictions.
Following that, \textcite{doskovc2021normal} investigated a number of semantic restrictions.

%% file: introduction/map.tex
\begin{tikzpicture}[above,sloped,shorten <=2mm,, shorten >=2mm]

\node[shape=rectangle,rounded corners,fill=lightgray] at (-3,-1) {$\Inf\Bc$};

\begin{scope}[every node/.style={minimum size=4mm}]
    \node (nothing) at (0,-0.4) {$\True$};
    \node (wmon) at (0,-1)     {$\WMon$};

    \node (mon) at (-1.5,-2.5)		  {$\Mon$}; 
    \node (dmon) at (1.5,-2.5)		  {$\DMon$}; 
    \node (bmon) at (0.0,-4)		  {$\BMon$}; 
    \node (smon) at (-1.5,-5.5)		  {$\SMon$}; 
    \node (dsmon) at (1.5,-5.5)		  {$\DSMon$}; 
    \node (bsmon) at (0.0,-7)		  {$\BSMon$}; 

    \draw (mon) -- (wmon);
    \draw (dmon) -- (wmon);
    \draw (bmon) -- (mon);
    \draw (bmon) -- (dmon);

    \draw (smon) -- (bmon);
    \draw (dsmon) -- (bmon);
    \draw (bsmon) -- (smon); 
    \draw (bsmon) -- (dsmon);
\end{scope}

\begin{pgfonlayer}{background}
    \node[line width=3pt, rounded corners, fit=(nothing) (wmon)] (Fitnothing) {};
    \foreach \x in {mon, dmon, bmon, smon, dsmon, bsmon}
        \node[line width=3pt, rounded corners, fit=(\x)] (Fit\x) {};

    \foreach \x in {nothing, mon, dmon, bmon, smon, dsmon, bsmon}
        \path[draw=lightgray,line width=3pt, rounded corners] (Fit\x.south east) -- (Fit\x.south west) -- (Fit\x.north west) -- (Fit\x.north east) -- cycle;
\end{pgfonlayer}

\end{tikzpicture}

%% file: introduction/preliminaries.tex
\input{introduction/preliminaries/math}

\section{Informants}
\input{introduction/preliminaries/informants}

\section{Interaction operators}
\input{introduction/preliminaries/interaction_operators}

\section{Learning restrictions}
\input{introduction/preliminaries/learning_restrictions}

\section{Properties of learning restrictions}
\input{introduction/preliminaries/restriction_properties}

%% file: introduction/preliminaries/math.tex
We refer the reader to the textbook by \textcite{rogers1987theory} for a more thorough introduction to computability theory.
We denote the set of natural numbers by $\N = \set{0, 1, 2, \dots}$ and the empty set by $\emptyset$.
For some set $S$, the cardinality of $S$ is written as $\cardinality S$.
By $\subseteq$, $\supseteq$, $\subsetneq$ and $\supsetneq$ we denote the relations subset, superset, proper subset and proper superset.
If two sets $A, B$ are incomparable, i.e.\ there are $x \in A \setminus B$ and $y \in B \setminus A$, we write $A \ncmp B$.
By $\min(A)$ and $\max(A)$, we denote the minimum and maximum of $A$.
By $\sup(A)$, we refer to the supremum of the set $A$, i.e.\ the smallest upper bound of $A$.
For convenience, we suppose that $\sup(\emptyset) = -\infty$.
For some set $S$, we denote the set of all sequences over $S$ by $\Seq(S)$.
By $\subSeq$ and $\subSeqNeq$ we denote the relations subsequence and proper subsequence.
The notation $\forall^\infty n \in \N$ means ``for all but finitely many $n \in \N$''.

If a partial function $f$ is undefined for some input $x$, we write $f(x)\divs$ or $f(x) = \bot$ and say $f$ diverges on input $x$.
Otherwise, $f$ converges on $x$, denoted by $f(x)\convs$.
If a function $f$ converges on all inputs $x \in \N$, we say that $f$ is total.
By $\partialFn$, $\partialCp$, $\totalFn$ and $\totalCp$, we denote the set of all functions, all computable functions, all total functions and all total computable functions respectively.
For some function $f$ and some number $n \in \N$, we denote the finite sequence of the first $n$ outputs of $f$ by $f[n]$.

We assume a numbering system $\varphi$ for the set of computable functions.
For each $f \in \partialCp$, there is $e \in \N$ with $\varphi_e = f$.
By $\phi$, we denote a Blum complexity measure for $\varphi$ \cite{blum1967machine}.
For example, for $e, x \in \N$, $\phi_e(x)$ could equal the number of steps the program coded by $e$ takes on input $x$.
By $\dom(f)$ and $\range(f)$, we refer to the domain and range of a function $f$.
A \emph{language} is a set of natural numbers.
By $\enum$, we denote the set of all recursively enumerable languages, i.e.\ all languages $L \subseteq \N$ for which there exists some $f \in \partialCp$ with $L = \dom(f)$.
To describe languages using natural numbers we use the $W$-indices system as our hypothesis space.
For all $e \in \N$, we define $W_e = \dom(\varphi_e)$, $e$ is called a \emph{label} for the language $W_e$.
Furthermore, for $e, t \in \N$, we define $W_e^t = \set{x \in \N \with x \leq t \land \phi_e(x) \leq t}$.
Notably, $W_e^t$ is finite and its construction is total and computable.
We define two numbers $a, b \in \N$ to be \emph{semantically equivalent}, denoted by $a \semequiv b$, if and only if $W_a = W_b$.
We use established theorems from computability theory, such as s-m-n, Kleene's Recursion Theorem (\KRT) \cite{kleene1952introduction} and the Operator Recursion Theorem (\ORT) \cite{case1994infinitary}.

%% file: introduction/preliminaries/informants.tex
Before we define what an informant is, we fix an interpretation for its output and introduce some useful vocabulary.
For any set $D \subseteq \N \times \bset$ we define \emph{positive information}, \emph{negative information} and \emph{outline} of $D$ as
\begin{align*}
    \pos(D)     &= \set{x \in \N \with (x, 1) \in D},\\
    \neg(D)     &= \set{x \in \N \with (x, 0) \in D},\\
    \outline(D) &= \pos(D) \cup \neg(D).
\end{align*}
For any finite or infinite sequence $\sigma$, we define
$$\content(\sigma) = \range(\sigma).$$
We define $\pos(\sigma)$, $\neg(\sigma)$ and $\outline(\sigma)$ analogously to the definition on sets.

A function $I: \N \rightarrow \N \times \bset$ is an \emph{informant} for a language $L \subseteq \N$ if $\pos(I) = L$ and $\neg(I) = \N \setminus L$.
By $\Inf$, we denote the set of all informants and for some language $L$ we define $\Inf(L)$ to be the set of informants for $L$.
An informant $I$ is \emph{canonical}, if and only if for all $x \in \N$, we have $I(x) = (x, 0)$ or $I(x) = (x, 1)$.
For all languages $L$, we define $\hat I_L$ to be the canonical informant for $L$.
Furthermore, we define $\InfCan$ to be set of all canonical informants.

%% file: introduction/preliminaries/interaction_operators.tex
According to an \emph{interaction operator}, the output of an informant $I$ may be presented to a learner $h$ in different formats.
For a learner $h \in \partialFn$ and informant $I \in \Inf$ we define the interaction operators $\G$ (Gold / full-information, \cite{gold1967language}), $\Psd$ (partially set-driven, \cite{schafer1984eingabeabhangigkeit}), $\Sd$ (set-driven, \cite{wexler1980formal}) and $\It$ (iterative, \cite{wexler1980formal}) that generate the \emph{hypothesis sequence} of a learner $h$ such that for all $i \in \N$ we have
\begin{align*}
    \G(h, I)(i)   &= h(I[i]), \\
    \Psd(h, I)(i) &= h(\content(I[i]), i), \\
    \Sd(h, I)(i)  &= h(\content(I[i])), \\
    \It(h, I)(i)  &= h(\It(h, I)(i-1), I[i]).
\end{align*}
Furthermore, we define the interaction operator $\CIt$ (confluently iterative, \cite{kotzing2016towards}), which is equivalent to $\It$, but also requires that the order and quantity in which the content is presented does not matter.

It is immediate that some operators pass strictly more information to the learner.
\Textcite{case2010strongly} manifest this notion by introducing a partial ordering of the interaction operators.
For two interaction operators $\beta, \beta'$ we define the relation $\trans$ such that
$$\beta \trans \beta' \iff [\forall h \in \partialFn \exists h' \in \partialFn \forall I \in \Inf: \beta(h, I) = \beta'(h', I)].$$
Intuitively, $\beta$-learners can be translated into $\beta'$-learners.
A slightly weakened version of this relation has been introduced by \textcite{kotzing2017normal} that only requires semantic equivalence of the hypothesis sequences.
For two interaction operators $\beta, \beta'$ we define the relation $\semTrans$ such that
$$\beta \semTrans \beta' \iff [\forall h \in \partialFn \exists h' \in \partialFn \forall I \in \Inf, n \in \N: \beta(h, I)(n) \semequiv \beta'(h', I)(n)].$$
Intuitively, $\beta$-learners can be translated into semantically equivalent $\beta'$-learners.
Additionally, if the relation goes both ways, we say $\beta \semTransEq \beta'$.

While $\CIt$ is standing out of line in our definition above, it is of great use for generalizing over interaction operators.
This is because it is a lower bound for all other interaction operators in the $\trans$ relation.
We have
\begin{gather*}
    \CIt \trans \Sd \trans \Psd \trans \G, \\
    \CIt \trans \It \trans \G, \\
    \It \semTransEq \G, \\
    \CIt \semTransEq \Sd.
\end{gather*}

%% file: introduction/preliminaries/learning_restrictions.tex
Now that we know how hypothesis sequences are generated, we define some restrictions on them.
Our restrictions are defined as predicates over the hypothesis sequence and the informant that generated that sequence.
When using the restrictions, we view them as predicates or sets containing all the elements satisfying the predicate interchangeably.
Let $\True$ be the predicate that is always true.
To start, we define two convergence criteria.

For a hypothesis sequence $p \in \partialFn$ generated from an informant $I \in \Inf$, we define the convergence criteria explanatory and behaviorally correct such that
\begin{align*}
    \Ex(p, I) &\iff [\exists q \forall^\infty n: p(n) = q \land W_q = \pos(I)], \\
    \Bc(p, I) &\iff [\forall^\infty n: W_{p(n)} = \pos(I)].
\end{align*}
Of course, $\Ex \subseteq \Bc$.
Next, we define the concept of consistency as first introduced by \textcite{angluin1980inductive}.
Consistency of a hypothesis $e \in \N$ with information $D~\subseteq~\N~\times~\bset$ is defined such that
$$\Cons(e, D) \iff [\pos(D) \subseteq W_e \land \neg(D) \cap W_e = \emptyset].$$
We define this analogously for languages described by their label $e$ and for other information formats, such as sequences.
Furthermore, we define consistency of a hypothesis sequence $p \in \partialFn$ with an informant $I \in \Inf$ such that
$$\Cons(p, I) \iff [\forall n \in \N: \Cons(p(n), I[n])].$$

Monotonicity restrictions are organized into three groups: strong monotonicity, monotonicity and weak monotonicity.
In each group, we have one variant that requires better generalizations, one that requires better specializations and one that is the conjunction of the two.
The generalization variants were all introduced first.
The dual (and combined) variants were all introduced by \textcite{lange1996monotonic}.
Strong monotonicity \cite{jantke1991monotonic} is the most strict of the there groups.
For all hypothesis sequences $p: \N \rightarrow \N$ and informants $I \in \Inf$, we define the restrictions strongly monotone, dual strongly monotone and combined strongly monotone, such that
\begin{align*}
    \SMon(p, I)  &\iff [\forall s, t: s \leq t \Implies W_{p(s)} \subseteq W_{p(t)}], \\
    \DSMon(p, I) &\iff [\forall s, t: s \leq t \Implies W_{p(s)} \supseteq W_{p(t)}], \\
    \BSMon(p, I) &\iff [\SMon(p, I) \land \DSMon(p, I)].
\end{align*}

Shortly after that, \textcite{wiehagen1990thesis} introduced the concept of (what we today call) monotonicity, which requires this behavior only on correct elements.
For all hypothesis sequences $p: \N \rightarrow \N$ and informants $I \in \Inf$, we define the restrictions monotone, dual monotone and combined monotone, such that
\begin{align*}
    \Mon(p, I)   &\iff [\forall s, t: s \leq t \Implies W_{p(s)} \cap \pos(I) \subseteq W_{p(t)} \cap \pos(I)], \\
    \DMon(p, I)  &\iff [\forall s, t: s \leq t \Implies \overline{W_{p(s)}} \cap \neg(I) \subseteq \overline{W_{p(t)}} \cap \neg(I)], \\
    \BMon(p, I)  &\iff [\Mon(p, I) \land \DMon(p, I)].
\end{align*}

Finally, \textcite{jantke1991monotonic} also introduced weak monotonicity where better generalizations are only required as long as new information does not contradict old hypotheses.
For all hypothesis sequences $p: \N \rightarrow \N$ and informants $I \in \Inf$, we define the restrictions weakly monotone, dual weakly monotone and combined weakly monotone, such that
\begin{align*}
    \WMon(p, I)  &\iff [\forall s, t: s \leq t \land \Cons(p(s), I[t]) \Implies W_{p(s)} \subseteq W_{p(t)}], \\
    \DWMon(p, I) &\iff [\forall s, t: s \leq t \land \Cons(p(s), I[t]) \Implies W_{p(s)} \supseteq W_{p(t)}], \\
    \BWMon(p, I) &\iff [\WMon(p, I) \land \DWMon(p, I)].
\end{align*}

Note that, $\SMon \subseteq \Mon \cap \WMon$.
This holds analogously for $\DSMon$ and $\BSMon$.
Furthermore, combined weak monotonicity is equivalent to \emph{semantic conservativeness} \cite{kotzing2017normal}.

The restriction of cautiousness was introduced by \textcite{osherson1982learning}.
For a learner to be cautious, we require it to never conjecture a proper subset of a previous hypothesis.
For explanatory learning, this poses a proper restriction as shown for example by \textcite{kotzing2016map} for text-learning and by \textcite{aschenbach2018learning} for informant-learning.
To investigate the restriction of cautiousness more closely, \textcite{kotzing2016map} introduced three new variants of cautiousness.
For all hypothesis sequences $p: \N \rightarrow \N$ and informants $I \in \Inf$, we define the restrictions cautious, target cautious, finitely cautious and infinitely cautious, such that
\begin{align*}
    \Caut(p, I)    &\iff [\forall s < t: \lnot(W_{p(s)} \supsetneq W_{p(t)})], \\
    \CautTar(p, I) &\iff [\forall s: \lnot (W_{p(s)} \supsetneq \pos(I))], \\
    \CautFin(p, I) &\iff [\forall s < t: W_{p(s)} \supsetneq W_{p(t)} \Implies W_{p(t)} \text{ is infinite}], \\
    \CautInf(p, I) &\iff [\forall s < t: W_{p(s)} \supsetneq W_{p(t)} \Implies W_{p(t)} \text{ is finite}].
\end{align*}

Note that, $\Caut = \CautFin \cap \CautInf$ and that $\SMon \subseteq \Caut \subseteq \CautTar$.

Learning restrictions can be combined.
For example, to require both consistency and behavioral correctness, we use the restriction $\Cons\Bc$.
We can require learning restrictions \emph{locally} or \emph{globally}.
A local restriction only has to be fulfilled when an informant for a target language is presented, global restriction must hold for any informant.
By $\Delta$, we denote the set of all learning restrictions (excluding convergence criteria) defined above, i.e.\
$$\Delta = \left\{ \parbox{0.7\textwidth}{$\Cons, \SMon, \DSMon, \BSMon, \Mon, \DMon, \BMon, $ $\WMon, \DWMon, \BWMon, \Caut, \CautTar, \CautFin, \CautInf$} \right\}.$$

In total, a learning setting consists of five elements:
The set of allowed learners $C$, the set of inputs used $P$, the interaction operator $\beta$, the global restriction $\alpha$ and the local restriction $\delta$.
A setting $S$ is thus defined as $S = (\alpha, P, C, \beta, \delta)$.
We refer to it as $C\tau(\alpha)P\beta\delta$, for example $\totalCp\tau(\Cons)\Inf\G\Bc$.
If $C = \partialCp$ or $\alpha = \True$, we omit writing the respective part.

Given some learning setting $S = (\alpha, P, C, \beta, \delta)$, we say that a learner $h$ $S$-learns the empty set, if $h \notin C$ or if there is some $I \in P$ for which $\alpha(\beta(h, I), I)$ is false.
Otherwise, it learns the set of languages
$$S(h) = \set{L \in \enum \with \forall I \in P(L): \delta(\beta(h, I), I)}.$$
By $[S]$, we denote the set of all $S$-learnable sets of languages (by any learner).

%% file: introduction/preliminaries/restriction_properties.tex
First, we introduce the concept of delayability \cite{kotzing2016map}.
Let $\simulatingFn$ denote the set of all non-decreasing, unbounded functions $\N \rightarrow \N$.
A learning restriction $\delta$ is called \emph{delayable}, if and only if for all informants $I, I' \in \Inf$ with $\content(I) = \content(I')$, hypothesis sequences $p$ and functions $s \in \simulatingFn$ the following implication holds.
If we have $(p, I) \in \delta$ and for all $n \in \N$ that $\content(I[s(n)]) \subseteq \content(I'[n])$, then $(p \circ s, I') \in \delta$.
Intuitively, a learning restriction is delayable, if we can skip or duplicate hypotheses in the hypothesis sequence, but the restriction still holds.
The second grouping we investigate is the one of \emph{semantic} restrictions.
The concept was first defined in \textcite{kotzing2017solution}, we follow the definitions by \textcite{kotzing2017normal} though.
For all $p \in \partialFn$, we fix the set
$$\Sem(p) = \set{p' \in \partialFn \with \forall i: (p(i)\convs \Iff p'(i)\convs) \land (p(i)\convs \Implies p(i) \semequiv p'(i))}.$$
A learning restriction $\delta$ is said to be \emph{semantic}, if for any sequence $p$ and informant $I$, we can conclude from $(p, I) \in \delta$ and $p' \in \Sem(p)$ that $(p', I) \in \delta$.
Intuitively, a semantic restriction is indifferent about semantically equivalent hypothesis sequences.
Out of all restrictions defined above, only $\Cons$ is \emph{not} delayable and only $\Ex$ is \emph{not} semantic.
Note that the combination of delayable restrictions is again delayable and the combination of semantic restrictions is again semantic.

%% file: core/plain.tex
In this chapter, we collect some theorems for delayable and for semantic restrictions.
In particular, we find that learning power is not dependent on the interaction operator (see \Cref{thm:interactionForDelaySem}).
At the end of the chapter, we use our findings about delayable restrictions to motivate the case for behaviorally correct learning by showing that it is strictly more powerful than explanatory learning (see \Cref{thm:bcexsep}).

\section{Delayable restrictions}

We first look at delayable restrictions and observe that correct identification from canonical informants suffices and that we can assume learners to be set-driven and total.
To do so, we combine two theorems by \textcite{aschenbach2018learning}.
As our new theorem builds on one of the proofs, we also include it here.

\begin{theorem}[\cite{aschenbach2018learning}]
    \label{thm:interactionForDelayable}
    For any delayable $\delta$, we have $[\InfCan\G\delta] = [\Inf\Sd\delta]$.
\end{theorem}
\begin{proof}
    By definition, $[\InfCan\G\delta] \supseteq [\Inf\Sd\delta]$.
    Let $h$ be a $\InfCan\G\delta$-learner and $\La = \InfCan\G\delta(h)$.
    We construct a learner $g$ that builds the longest possible prefix of a canonical informant from the information it has.
    It then passes this prefix to $h$.
    Recall that for any language $L$, $\hat I_L$ is the canonical informant for $L$.

    For any finite set $D \subseteq \N \times \bset$, we define the length of the prefix as \linebreak $\ell(D) = \max\set{n \in \N \with \forall i < n: i \in \outline(D)}$.
    Next, the prefix itself is defined as $c(D) = \hat I_{\pos(D)}[\ell(D)]$.
    Finally, our prediction is $g(D) = h(c(D))$.

    Let $L \in \La$ and let $I \in \Inf(L)$ be any informant for $L$.
    In order to apply delayability we need a simulating function $s \in \simulatingFn$.
    We pick $s: n \mapsto \ell(\content(I[n]))$.
    By definition of $\ell$, $s$ is non-decreasing.
    Since $I$ is an informant, $s$ must also be unbounded.
    Therefore, $s \in \simulatingFn$.

    Let $n \in \N$.
    By definition of $c$ and $s$, we have $\hat I_L[s(n)] = \hat I_L[\ell(\content(I[n]))] = c(\content(I[n]))$.
    Thus, $\content(\hat I_L[s(n)]) \subseteq \content(I[n])$ and
    $$g(\content(I[n])) = h(c(\content(I[n]))) = h(\hat I_L[s(n)]).$$
    Therefore, $\Sd(g, I) = \G(h, \hat I_L) \circ s$.
    As $\delta$ is delayable and $(\G(h, \hat I_L), \hat I_L) \in \delta$, we can conclude that $(\Sd(g, I), I) = (\G(h, \hat I_L) \circ s, I) \in \delta$.
    Hence, $\La$ is $\Inf\G\delta$-learnable.
\end{proof}

\begin{theorem}
    \label{thm:totalInteractionForDelayable}
    For any delayable $\delta$, we have $[\InfCan\G\delta] = [\totalCp\Inf\Sd\delta]$.
\end{theorem}
\begin{proof}
    \Textcite{aschenbach2018learning} showed that $[\totalCp\Inf\G\delta] = [\Inf\G\delta]$.
    By \Cref{thm:interactionForDelayable}, we have $[\InfCan\G\delta] = [\Inf\G\delta]$ and thus $[\InfCan\G\delta] = [\totalCp\InfCan\G\delta]$.
    This means, that we can assume the $\G$-learner picked in the proof of \Cref{thm:interactionForDelayable} to be total, making the constructed $\Sd$-learner total as well.
\end{proof}

\section{Semantic restrictions}

With delayable restrictions, we get equivalent learning capabilities for interaction operators until $\Sd$.
To include iterative learners, we need to look at semantic restrictions.
\Textcite{kotzing2017normal} showed that $\G \semTransEq \It$ and $\Sd \semTransEq \CIt$, allowing us to fill the gap.
Firstly, they showed that for semantic $\delta$ and for any interaction operator $\beta$, we have $[\totalCp\Txt\beta\delta] = [\Txt\beta\delta]$.
The same proof they employed may be used to show that this also holds for informants.

\begin{theorem}
    \label{thm:semanticGivesTotal}
    For any interaction operator $\beta$ with $\CIt \trans \beta \trans \G$ and semantic restriction $\delta$ holds $[\totalCp\Inf\beta\delta] = [\Inf\beta\delta]$.
\end{theorem}

We can combine this with our observations for delayable learning restrictions, to get the final result.

\begin{theorem}
    \label{thm:interactionForDelaySem}
    For all learning restrictions $\delta$ that are delayable and semantic, for all interaction operators $\beta$ with $\CIt \trans \beta \trans \G$ we have $[\totalCp\Inf\beta\delta] = [\InfCan\G\delta]$.
\end{theorem}
\begin{proof}
    One inclusion is trivial.
    Since $\delta$ is delayable, \Cref{thm:interactionForDelayable} gives us that $[\InfCan\G\delta] = [\Inf\Sd\delta]$.
    We can use that $\delta$ is semantic and $\CIt \semTransEq \Sd$ to get $[\Inf\Sd\delta] = [\Inf\CIt\delta]$.
    By $\CIt \trans \beta$ we get $[\Inf\CIt\delta] \subseteq [\Inf\beta\delta]$.
    Finally, \Cref{thm:semanticGivesTotal} gives us $[\Inf\beta\delta] = [\totalCp\Inf\beta\delta]$.
\end{proof}

\begin{corollary}
    \label{thm:interactionForBc}
    For all $\CIt \trans \beta \trans \G$ and all learning restrictions $\delta \in \Delta \setminus \set{\Cons}$ we have $[\Inf\beta\delta\Bc] = [\InfCan\G\delta\Bc]$.
\end{corollary}

\section{Separation from explanatory learning}

\Textcite{kotzing2016towards} observed that $[\Txt\Sd\SMon\Bc] \setminus [\Txt\G\Ex] \neq \emptyset$.
We modify this proof to work on informants, giving us a non-constructive proof that is more direct than the one by \textcite{aschenbach2018learning}.

\begin{theorem}[\cite{aschenbach2018learning}]
    \label{thm:bcexsep}
    We have $[\Inf\Sd\SMon\Bc] \setminus [\Inf\G\Ex] \neq \emptyset$.
\end{theorem}
\begin{proof}
    We consider the following $\Sd$-learner $h$, which maps a finite set $D \subseteq \N \times \bset$ to the hypothesis
    $$W_{h(D)} = \begin{cases}
        \emptyset,         & \text{if } \pos(D) = \emptyset, \\
        W_{\max(\pos(D))}, & \text{otherwise}.
    \end{cases}$$

    Let $\La = \Inf\Sd\SMon\Bc(h)$.
    By \Cref{thm:totalInteractionForDelayable} we have $[\Inf\G\Ex] = [\totalCp\Inf\Sd\Ex]$.
    By way of contradiction, assume that $\La \in [\totalCp\Inf\Sd\Ex]$ as witnessed by some learner $g \in \totalCp$.
    Using $\ORT$, we get a computable sequence $(D_i)_{i \in \N}$ of sets as well as a total recursive function $p \in \totalCp$ strongly monotone satisfying for all finite $D \subseteq \N \times \bset$ and $b \in \bset, i, t \in \N$
    \footnote{For convenience, we suppose that $\sup(\emptyset) = -\infty$.}
    \begin{align*}
        p(D, b) &> \sup(\outline(D)),\\
        W_{p(D, b)} &= \set{p(D, b)} \cup \bigcup_{j \in \N; D_j \convs} \pos(D_j),\\
        \suc(D, b, t) &= \content\left(\hat I_{\pos(D) \cup \set{p(D, b)}}[p(D, b) + t]\right) ,\\
        D_0 &= \emptyset,\\
        D_{i+1} &= \begin{cases}
            \suc(D_i, b, t), &\text{if } \exists b \in \bset, t \in \N: g(D_i) \neq g(\suc(D_i, 0, t)),\\
            \bot,            &\text{otherwise}.
        \end{cases}
    \end{align*}
    In every iteration of the $(D_i)_{i \in \N}$ sequence, the canonical informant describes a finite language that is one element larger than in the previous iteration.
    By our first restriction, the previous set did not include any information about the newly added element and thus, the new set does not contradict the previous one.
    The $t$ parameter modifies how much additional negative information is included.

    As each $D_i$ matches the $\content$ of a prefix of a canonical informant, $D_i$ is included in the input sequence for an $\Sd$-learner that would result from a canonical informant for a superset of $D_i$.
    Consider the language $L = \bigcup_{i \in \N; D_i\convs}\pos(D_i)$.

    \paragraph{Case 1: $L$ is infinite.}
    We have for any $p(D, b) \in L$ that
    $$W_{p(D, b)} = \set{p(D, b)} \cup \bigcup_{i \in \N} \pos(D_i) = \set{p(D, b)} \cup L = L.$$
    As $h$ only outputs numbers that are contained in $L$, $L \in \La$.
    On the other hand, $g$ cannot learn $L$ from the canonical informant for $L$ as it makes infinitely many mind changes by the construction of $(D_i)_{i \in \N}$.

    \paragraph{Case 2: $L$ is finite.}
    That means, from one point on the sets $D_i$ are undefined.
    Let $D_k$ be the last set defined.
    As we have for $b \in \bset$ that
    $$\sup(L) = \sup(\pos(D_k)) \leq \sup(\outline(D_k)) < p(D_k, b),$$ we know $p(D_k, b) \notin L$.
    Consider the following two proper supersets of $L$:
    \begin{align*}
        L_0 &= \set{p(D_k, 0)} \cup L,\\
        L_1 &= \set{p(D_k, 1)} \cup L.
    \end{align*}
    For $b \in \bset$, let $I \in \Inf(L_b)$ be an informant for $L_b$.
    When $I$ shows its first positive information, $h$ conjectures the set $L$ until the positive information about $\max(L_b) = p(D_k, b)$ is presented.
    Now $h$ switches to its final and correct guess $W_{p(D_k, b)} = L_b$.
    This sequence of hypotheses fulfills $\SMon$, so $L_0, L_1 \in \La$.

    From the definition of the sequence $(D_i)_{i \in \N}$, we know that $g$ converges on the canonical informants for $L_0$ and $L_1$ to the same hypothesis, namely $g(D_k)$.
    Hence, $g$ cannot learn both languages $L_0$ and $L_1$ as they are different, a contradiction.
\end{proof}

%% file: core/cautiousness.tex
In this section, we observe that, for learning from informants, requiring learners to be cautious poses a proper restriction.
In particular, we find that all three reduced variants of cautiousness properly restrict learners.
This is in contrast to what has been found for text-learning by \textcite{kotzing2016map} where $\CautInf$ does not lessen learning power.

\begin{theorem}
    \label{thm:cauttarandinfproper}
    We have $[\Inf\G\Mon\Bc] \setminus [\Inf\G\CautTar\Bc] \neq \emptyset$ and \linebreak $[\Inf\G\Mon\Bc] \setminus [\Inf\G\CautInf\Bc] \neq \emptyset$.
\end{theorem}
\begin{proof}
    This proof is analogous to the separation of $\Caut$ from $\Mon$ by \textcite{aschenbach2018learning}.
    Consider $\La = \set{\N \setminus D \with D \subseteq \N \text{ and $D$ is finite}}$.
    It is easy to see that $\La$ is $\Inf\G\Mon\Bc$-learnable by the learner $h$ that maps a finite sequence $\sigma \in \Seq(\N \times \bset)$ to the hypothesis $W_{h(\sigma)} = \N \setminus \neg(\sigma)$.

    Let $g$ be a $\Inf\G\Bc$-learner for $\La$.
    As $\N \in \La$, there is $n_0$ such that for all $n \geq n_0$ we have $W_{g(\hat I_\N[n_0])} = \N$.
    Let $L = \N \setminus \set{n_0 + 1}$.
    Then, $\hat I_L[n_0] = \hat I_\N[n_0]$ and thus $W_{g(\hat I_L[n_0])} = \N \supsetneq L$.
    As $L = \pos(\hat I_L)$ and as $L$ is infinite, $g$ cannot learn $\La$ while preserving $\CautTar$ or $\CautInf$.
\end{proof}

\begin{corollary}
    \label{thm:cautproper}
    We have $[\Inf\G\Mon\Bc] \setminus [\Inf\G\Caut\Bc] \neq \emptyset$.
\end{corollary}

\begin{theorem}
    We have $[\Inf\G\Bc] \setminus [\Inf\G\CautFin\Bc] \neq \emptyset$.
\end{theorem}
\begin{proof}
    Consider $\La = \set\N \cup \set{D \subseteq \N \with D \text{ is finite}}$.
    $\La$ is $\Inf\G\Bc$-learnable by the leaner $h$ that maps a finite sequence $\sigma \in \Seq(\N \times \bset)$ to the hypothesis
    \begin{align*}
        W_{h(\sigma)}  &= \begin{cases}
            \N,           & \text{if } \neg(\sigma) = \emptyset, \\
            \pos(\sigma), & \text{otherwise}.
        \end{cases}
    \end{align*}
    The proof that $\La \notin [\Inf\G\CautFin\Bc]$ is analogous to \Cref{thm:cauttarandinfproper}, with the modification that we return to the finite set of all the positive information shown so far after the learner conjectures $\N$.
\end{proof}

%% file: core/monotonicity.tex
In this section, we investigate the nine variants of monotonicity.
We start by finding that weak monotonicity does not properly restrict learners.
This is in line with what has been found for explanatory learning.
For example, \textcite{aschenbach2018learning} showed that $[\Inf\G\Ex] = [\Inf\G\WMon\Ex]$.
In \Cref{thm:conswmonistrue} we present a new construction for a weakly monotonic learner that is also globally consistent.

Afterwards, we compare the classic and dual versions of monotonicity and strong monotonicity.
We find that in both cases, the sets of learnable language classes are incomparable (see \Cref{thm:sdmonncmp} and \Cref{thm:mondualmonsep}).
Then, we separate the classic versions from one another (see \Cref{thm:sepmonfromtrue} and \Cref{thm:sepMonSMon}) and show that strong monotonicity and dual strong monotonicity imply combined monotonicity (see \Cref{thm:smonanddualimplycombinedmon}).
For a map of all relations, see \Cref{fig:monmap}.

\begin{theorem}
    \label{thm:conswmonistrue}
    We have $[\Inf\G\Bc] = [\tau(\Cons)\Inf\G\WMon\Bc]$.
\end{theorem}
\begin{proof}
    One inclusion holds by definition.
    For the other, let $h$ be a $\G$-learner and $\La \subseteq \Inf\G\Bc(h)$.
    By \Cref{thm:totalInteractionForDelayable}, we can assume that $h$ is total.
    Consider the learner $g$ that maps a finite sequence $\sigma \in \Seq(\N \times \bset)$ to the hypothesis
    \begin{align*}
        E(\sigma)     &= \set{h(\sigma)} \cup \set{g(\tau) \with \tau \subSeqNeq \sigma}, \\
        W_{g(\sigma)} &= \pos(\sigma) \cup \bigcup_{e \in E(\sigma)} \bigcup_{t \in \N} \begin{cases}
            W_e^t,     & \text{if } \Cons(W_e^t, \sigma), \\
            \emptyset, & \text{otherwise}.
        \end{cases}
    \end{align*}
    Intuitively, we use $h$'s hypothesis and, in order to remain weakly monotonic, collect all our previous hypotheses which are consistent with our current information.

    Clearly, $g$ is globally consistent.
    Let $L \in \La$ and $I \in \Inf(L)$.
    To show weak monotonicity, let $i, j \in \N$ with $i < j$ and suppose $\Cons(g(I[i]), I[j])$.
    Then, for all but finitely many $t \in \N$, we have $\Cons(W_{g(I[i])}^t, I[j])$ and, as $g(I[i]) \in E(I[j])$, we have $W_{g(I[i])} \subseteq W_{g(I[j])}$.

    Finally, we show that $g$ $\Bc$-identifies $L$.
    Let $n_0 \in \N$ such that for all $n \geq n_0$ we have $W_{h(I[n])} = L$.
    As we include $h$'s hypotheses, we have for all $n \geq n_0$ that $L \subseteq W_{g(I[n])}$.
    It remains to show that the possibly infinitely many wrong elements included in $W_{g(I[n_0])}$ are sorted out at some point.
    To do so, we first observe that there is a point $n_2$ at which all wrong hypotheses among the first $n_0$ hypotheses are sorted out.
    Then, assuming the hypothesis is still incorrect, we show that when new negative information is shown, there is some wrong element that is included in all remaining wrong hypotheses.
    As the enumeration of the wrong previous hypotheses stops once this element is found, only finitely many wrong elements can remain.
    Those are removed once they appear in the negative information of $I$.
    We proceed with the formal proof.

    Since we have, for all $n \geq n_0$, that $W_{h(I[n])} = L$, no new wrong elements are introduced after the first $n_0$ hypotheses.
    Thus for all $n \geq n_0$ we have $W_{g(I[n])}~\supseteq~W_{g(I[n+1])}$.
    Firstly, let $n_1 \geq n_0$ be such that for all $n < n_0$ with $W_{g(I[n])} \not\subseteq L$ there is an element contradicting the consistency of $g(I[n])$ in $\neg(I[n_1])$.
    As enumeration of these hypotheses stops once the wrong elements are witnessed, only finitely many wrong elements from guesses before $n_0$ remain in hypotheses $W_{g(I[n])}$ for $n \geq n_1$.
    Let $n_2$ be such that all of those are contained in $\neg(I[n_2])$.
    If $W_{g(I[n_2])} = L$, we are done.
    Otherwise, it remains to show that the wrong elements from $W_{g(I[n])}$ for $n_1 \leq n < n_2$ that are still included in $W_{g(I[n_2])}$ are eventually removed.

    Since $I$ presents each element from $\overline L$ at some point, there is $n_3 > n_2$ with $W_{g(I[n_3-1])} \supsetneq W_{g(I[n_3])}$.
    As $g$ is weakly monotonic, this implies that \linebreak $g(I[n_3-1])$ is not consistent with $I[n_3]$.
    Since $\pos(I[n_3]) \subseteq L \subseteq W_{g(I[n_3-1])}$, there is $x \in \neg(I[n_3]) \cap W_{g(I[n_3-1])}$.
    Since we have for all $n$ with $n_0 \leq n$ that $W_{g(I[n])} \supseteq W_{g(I[n+1])}$, we have for all $n$ with $n_0 \leq n < n_3$ that $x \in W_{g(I[n])}$.
    Let
    $$t_x = \max\set{\min\set{t \in \N \with x \in W_{g(I[n])}^t} \with n_0 \leq n < n_3}.$$
    Then, for $t \geq t_x$ and $\tau \subSeqNeq I[n_3]$, all $W_{g(\tau)}^t$ are inconsistent with $I[n_3]$.
    Hence, $W_{g(I[n_3])}$ consists of only $W_{h(I[n_3])} = L$ and at most $t_x$ many wrong elements from previous guesses.
    Let $n_4 > n_3$ such that all of those wrong elements are included in $\neg(I[n_4])$, then for all $n \geq n_4$ we have $W_{g(I[n])} = L$.
    Therefore, $g$ identifies $L$ and $\Bc$-learns $\La$.
\end{proof}

\section{Separation of classic and dual variants}

In this section, we show that for both monotonicity and strong monotonicity, the set of learnable languages classes by the classic and dual variants are incomparable.
This is a key observation, as it suffices to conclude the complete map.
All separations presented in this section are topological and transferred from analogous proofs by \textcite{lange1994characterization} where they occurred in the setting of explanatory learning of indexed families.

\begin{lemma}
    \label{thm:smonMinusDsmon}
    We have $[\Inf\G\SMon\Bc] \setminus [\Inf\G\DSMon\Bc] \neq \emptyset$.
\end{lemma}
\begin{proof}
    Consider $\La = \set{D \subseteq \N \with D \text{ finite}}$.
    The set can be learned by the strong monotonic learner $h = \pos$.
    It cannot be learned by a dual strong monotonic learner.
    Intuitively, to infer a language $L \in \La$, a learner has to conjecture a label for $L$ at some point.
    When making this guess, there is some element $x \notin L$ for which no information has been given so far.
    Under dual strong monotonicity, the learner cannot include $x$ in its later hypotheses.
    Thus, it cannot learn $L \cup \set x \in \La$.
\end{proof}

\begin{lemma}
    \label{thm:dsmonMinusSmon}
    We have $[\Inf\G\DSMon\Bc] \setminus [\Inf\G\SMon\Bc] \neq \emptyset$.
\end{lemma}
\begin{proof}
    Consider $\La = \set\N \cup \set{\set{0, 1, \dots, n} \with n \in \N}$.
    The learner $h$, that maps a finite sequence $\sigma \in \Seq(\N \times \bset)$ to the hypothesis
    $$W_{h(\sigma)} = \begin{cases}
        \N,                                      &\text{if } \neg(\sigma) = \emptyset,\\
        \set{0, 1, \dots, \min(\neg(\sigma))-1}, &\text{otherwise}
    \end{cases}$$
    learns $\La$ dual strong monotonically.

    $\La$ cannot be learned strong monotonically though.
    Suppose there is a learner $g$ that $\Inf\SMon\Bc$-learns $\La$.
    Let $n \in \N$ be such that $W_{g(\hat I_\N[n])} = \N$.
    Then $g$ cannot infer $\set{0, 1, \dots, n+1}$ from its canonical informant, because it guesses a program for $\N$ after seeing the first $n$ elements.
    Hence, $\La \notin [\Inf\SMon\Bc]$.
\end{proof}

\begin{theorem}
    \label{thm:sdmonncmp}
    We have $[\Inf\SMon\Bc] \ncmp [\Inf\DSMon\Bc]$.
\end{theorem}
\begin{proof}
    By $\Cref{thm:smonMinusDsmon}$ and $\Cref{thm:dsmonMinusSmon}$.
\end{proof}

\begin{lemma}
    \label{thm:monMinusDmon}
    We have $[\Inf\G\Mon\Bc] \setminus [\Inf\G\DMon\Bc] \neq \emptyset$.
\end{lemma}
\begin{proof}
    For all $i \in \N$, let $a_i = 3i$, $b_i = 3i + 1$ and $c_i = 3i + 2$.
    For all $n, m \in \N$ with $n < m$, consider the languages
    \begin{align*}
        X        &= \set{a_i \with i \in \N}, \\
        Y_n      &= \set{a_i \with i \leq n} \cup \set{b_i \with n < i}, \\
        Z_{n, m} &= \set{a_i \with i \leq n} \cup \set{b_i \with n < i \leq m} \cup \set{c_m}.
    \end{align*}
    Let $\La = \set X \cup \set{Y_n \with n \in \N} \cup \set{Z_{n,m} \with n < m}$.
    Intuitively, the languages in $\La$ describe the contents of streams that start out listing $a$s, then switch to $b$s and then end with a $c$.
    However, these streams may also continue listing $a$s or $b$s indefinitely.

    Consider the learner $h$ that maps a finite sequence $\sigma \in \Seq(\N \times \bset)$ to the hypothesis
    $$W_{h(\sigma)} = \begin{cases}
        Z_{n,m}, &\text{if } \exists n, m \in \N: \set{a_n, b_{n+1}, c_m} \subseteq \pos(\sigma), \\
        Y_n,     &\text{else if } \exists n \in \N: \set{a_n, b_{n+1}} \subseteq \pos(\sigma),    \\
        X,       &\text{otherwise}.
    \end{cases}$$
    Intuitively, $h$ conjectures $X$ until the boundary between the $a$s and $b$s is included in the information.
    Then, it conjectures the according $Y_n$, until a $c_m$ is presented.
    This hypothesis sequence is monotonic, $\La \in [\Inf\G\Mon\Bc]$.

    Suppose that $\La$ is $\Inf\G\DMon\Bc$-learnable as witnessed by some learner $g$.
    For some $n, m \in \N$, we let $g$ infer $Z_{n,m}$, but force it to conjecture $X$ and $Y_n$ before $Z_{n,m}$.
    To break dual monotonicity, we show that there is an element $b \notin Z_{n,m}$, which will be included in $Y_n$, but not in $X$.

    Let $I_X \in \Inf(X)$.
    Then there is $n_X \in \N$ such that $W_{g(I_X[n_X])} = X$.
    Let $n \in \N$ be such that for all $i \geq n$ we have $a_i, b_i, c_i \notin \outline(I_X[n_X])$.
    This means, that $g$ conjectures $X$ without knowing whether $\pos(I_X)$ really describes an infinite stream of $a$s, or whether it may change to list $b$s instead.
    Let $I_Y$ be an informant for $Y_n$ such that $I_X[n_X] = I_Y[n_X]$.
    Such an informant exists, because by definition of $n$, we know $\pos(I_X[n_X]) \subseteq \set{a_i \with i < n}$ and $\neg(I_X[n_X]) \cap \set{b_i \with i \in \N} \subseteq \set{b_i \with i < n}$.
    Let $n_Y \in \N$ with $n_Y > n_X$ be such that $W_{g(I_Y[n_Y])} = Y_n$.
    Similarly to $n$, let $m \in \N$ be such that $m > n$ and for all $i \geq m$ we have $a_i, b_i, c_i \notin \outline(I_Y[n_Y])$ and let $I_Z \in \Inf(Z_{n,m})$ with $I_Y[n_Y] = I_Z[n_Y]$.
    Given $I_Z$, $g$ first conjectures $X$, then $Y$ and finally $Z$.
    We have $b_{m+1} \notin X$ and $b \in Y$.
    As $b_{m+1} \notin Z$, dual monotonicity is violated, a contradiction.
\end{proof}

\begin{lemma}
    \label{thm:dmonMinusMon}
    We have $[\Inf\G\DMon\Bc] \setminus [\Inf\G\Mon\Bc] \neq \emptyset$.
\end{lemma}
\begin{proof}
    For all $n, m \in \N$ with $n < m$, consider the languages
    \begin{align*}
        X &= 2\N, \\
        Y_n &= \set{2n + 1} \cup \set{2i \with i \leq n}, \\
        Z_{n, m} &= Y_n \cup \set{2m}.
    \end{align*}
    Let $\La = \set X \cup \set{Y_n \with n \in \N} \cup \set{Z_{n,m} \with n,m \in \N, n < m}$.
    Consider the learner $h$ that maps a finite sequence $\sigma \in \Seq(\N \times \bset)$ to the hypothesis
    $$W_{h(\sigma)} = \begin{cases}
        Z_{n, m}, & \text{if } \exists n,m \in \N: n < m \land \set{2n, 2n+1, 2m} \subseteq \pos(\sigma), \\
        Y_n,      & \text{else if } \exists n \in \N: \set{2n, 2n+1} \subseteq \pos(\sigma),              \\
        X,        & \text{otherwise}.
    \end{cases}$$
    Intuitively, $h$ conjectures $X$ until for some $n \in \N$, we find $2n+1$ in the positive data, suggesting that the target language is $Y_n$.
    Then, if another even number $2m > 2n+1$ is found, $h$ switches to $Z_{n,m}$.
    This hypothesis sequence is dual monotonic, so $\La \in [\Inf\G\DMon\Bc]$.

    Suppose that $\La$ is $\Inf\G\Mon\Bc$-learnable as witnessed by some learner $g$.
    For some $n, m \in \N$, we let $g$ infer $Z_{n,m}$, but force it to conjecture $X$ and $Y_n$ before $Z_{n,m}$.
    To break monotonicity, we show that the number $2m \in Z_{n,m}$ will be included in $X$, but not in $Y_n$.

    Let $n_X \in \N$ such that $W_{g(\hat I_X[n_X])} = X$.
    Let $n = n_X + 1$ and $Y = Y_n$.
    Now, let $n_Y > n_X$ such that $W_{g(\hat I_Y[n_Y])} = Y$.
    Note that $\hat I_X[n_X] = \hat I_Y[n_X]$, as both include all even numbers and exclude all odd numbers up to $2n_X$.
    Finally, let $m = n_Y + 1$ and $Z = Z_{n,m}$.
    Let $n_Z > n_Y$ such that $W_{g(\hat I_Z[n_Z])} = Z$.
    Again, $\hat I_Y[n_Y] = \hat I_Z[n_Y]$.
    This means when inferring $Z \in \La$ from its canonical informant, $g$ conjectures $X$, then $Y$ and then $Z$.
    We have $2m \in X$ and $2m \notin Y$.
    As $2m \in Z$, monotonicity is violated, a contradiction.
\end{proof}

\begin{theorem}
    \label{thm:mondualmonsep}
    We have $[\Inf\G\Mon\Bc] \ncmp [\Inf\G\DMon\Bc]$.
\end{theorem}
\begin{proof}
    By \Cref{thm:monMinusDmon} and \Cref{thm:dmonMinusMon}.
\end{proof}

\section{Completing the picture of monotonic constraints}

In this section, we collect all the other relations between monotonic learning restrictions.
In particular, we show that both variants of strong monotonicity imply combined monotonicity.
All other theorems are implied by our previous separations.

\begin{corollary}
    \label{thm:sepmonfromtrue}
    We have $[\Inf\G\Mon\Bc] \subsetneq [\Inf\G\Bc]$ as well as \linebreak $[\Inf\G\DMon\Bc] \subsetneq [\Inf\G\Bc]$.
\end{corollary}
\begin{proof}
    By definition, both $[\Inf\G\Mon\Bc]$ and $[\Inf\G\DMon\Bc]$ are subsets of \linebreak $[\Inf\G\Bc]$.
    As they are incomparable by \Cref{thm:mondualmonsep}, neither of them can be equal to $[\Inf\G\Bc]$ though.
    Therefore, both are proper subsets.
\end{proof}

\begin{corollary}
    We have $[\Inf\G\BMon\Bc] \subsetneq [\Inf\G\Mon\Bc]$ as well as \linebreak $[\Inf\G\BMon\Bc]~\subsetneq~[\Inf\G\DMon\Bc]$.
\end{corollary}
\begin{proof}
    The reasoning is the same as for \Cref{thm:sepmonfromtrue}.
\end{proof}

\begin{corollary}
    \label{thm:bmonpropersup}
    We have $[\Inf\G\BSMon\Bc] \subsetneq [\Inf\G\SMon\Bc]$ as well as \linebreak $[\Inf\G\BSMon\Bc] \subsetneq [\Inf\G\DSMon\Bc]$.
\end{corollary}
\begin{proof}
    The reasoning is the same as for \Cref{thm:sepmonfromtrue}.
\end{proof}

\begin{theorem}
    \label{thm:smonanddualimplycombinedmon}
    We have $[\Inf\G\SMon\Bc] \subsetneq [\Inf\G\BMon\Bc]$ as well as \linebreak $[\Inf\G\DSMon\Bc] \subsetneq [\Inf\G\BMon\Bc]$.
\end{theorem}
\begin{proof}
    Consider a $\G$-learner $h$ and $\La = \Inf\G\SMon\Bc(h)$.
    Let $L \in \La$ and $I \in \Inf(L)$.
    Since $h$ is strongly monotonic, we have for all $t \in \N$ that $W_{h(I[t])} \subseteq L$.
    We get $\neg(I[t]) \subseteq \overline L \subseteq \overline{W_{h(I[t])}}$ and thus $\overline{W_{h(I[t])}} \cap \neg(I[t]) = \neg(I[t])$.
    For $s, t \in \N$ with $s \leq t$ we have
    $$\overline{W_{h(I[s])}} \cap \neg(I[s]) = \neg(I[s]) \subseteq \neg(I[t]) = \overline{W_{h(I[t])}} \cap \neg(I[t]).$$
    Hence, $\SMon$ not only implies $\Mon$, but also $\DMon$ and therefore $\BMon$.
    For a dual strong monotonic learner $h$, we can show that for all $t \in \N$ we have \linebreak $W_{h(I[t])} \cap \pos(I[t]) = \pos(I[t])$ and consequently that $h$ is $\BMon$.
    In conclusion, we know that both $[\Inf\G\SMon\Bc]$ and $[\Inf\G\DSMon\Bc]$ are subsets of $[\Inf\G\BMon\Bc]$.
    Since $[\Inf\G\SMon\Bc]$ and $[\Inf\G\DSMon\Bc]$ are incomparable by \Cref{thm:sdmonncmp}, neither of them can be equal to $[\Inf\G\BMon\Bc]$, so they are both proper subsets.
\end{proof}

\begin{corollary}
    \label{thm:sepMonSMon}
    We have $[\Inf\Mon\Bc] \setminus [\Inf\SMon\Bc] \neq \emptyset$.
\end{corollary}
\begin{proof}
    As $\SMon \subseteq \Caut$, this is a direct consequence of \Cref{thm:cautproper}.
\end{proof}

\begin{corollary}
    We have $[\Inf\G\DSMon\Bc] \subsetneq [\Inf\G\DMon\Bc]$.
\end{corollary}
\begin{proof}
    This is a direct consequence of \Cref{thm:smonanddualimplycombinedmon} and \Cref{thm:bmonpropersup}.
\end{proof}

%% file: core/consistency.tex
In this section, we observe that, on its own, consistency does not restrict $\Inf\Bc$-learners, as the seen data can easily be patched into hypotheses (see \Cref{thm:plainglobalcons}).
Using this approach, we can also preserve all variants of monotonicity and strong monotonicity (see \Cref{thm:patchPreservesBDSMon}).
For weak monotonicity, we already observed that global consistency can be assumed in \Cref{thm:conswmonistrue}.
We conclude the chapter with \Cref{thm:makedualwmoncons} where we employ \emph{poisoning} to add global consistency to a dual weakly monotonic learner.

\section{Patching learners}

When patching hypotheses with the information seen so far, we can achieve consistency, while maintaining correct hypotheses.
First, we make some general observations about patched hypotheses and then use those to add consistency to a set-driven $\Bc$-learner.
Afterwards, we conclude that this suffices to show that all learners without further restrictions can be assumed to be globally consistent.

\begin{definition}
    We define the function $\patch \in \totalCp$ such that for all $e \in \N$ and finite sets $D \subseteq \N \times \bset$ we have
    $$W_{\patch(e, D)} = (W_e \cup \pos(D)) \setminus \neg(D).$$
\end{definition}

\begin{lemma}
    \label{thm:patchMakesCons}
    For all informants $I \in \Inf$ and numbers $e, t \in \N$, we have $\Cons(\patch(e, \content(I[t])), I[t])$.
\end{lemma}
\begin{proof}
    Let $D = I[t]$.
    As $I$ is an informant, we have $\pos(D) \cap \neg(D) = \emptyset$.
    Hence,
    \begin{align*}
        \pos(D) & \subseteq (W_e \setminus \neg(D)) \cup \pos(D) \\
                & = (W_e \cup \pos(D)) \setminus \neg(D)         \\
                & = W_{\patch(e, D)}.
    \end{align*}
    Furthermore, $W_{\patch(e, D)} \cap \neg(D) = ((W_e \cup \pos(D)) \setminus \neg(D)) \cap \neg(D) = \emptyset$.
\end{proof}

\begin{lemma}
    \label{thm:patchMaintainsCorrect}
    For numbers $e \in \N$ and informants $I \in \Inf(W_e)$ we have for all $t$ that $\patch(e, \content(I[t])) \semequiv e$.
\end{lemma}
\begin{proof}
    Let $t \in \N$ and $D = I[t]$.
    As $I$ is an informant, we have $\pos(D) \subseteq \pos(I)$ and $\neg(D) \cap \pos(I) = \emptyset$.
    Then we have
    \begin{align*}
        W_{\patch(e, D)} & = (W_e \cup \pos(D)) \setminus \neg(D)     \\
                         & = (\pos(I) \cup \pos(D)) \setminus \neg(D) \\
                         & = \pos(I)                                  \\
                         & = W_e.
    \end{align*}
\end{proof}

\begin{theorem}
    \label{thm:plainglobalcons}
    We have $[\tau(\Cons)\Inf\Sd\Bc] = [\Inf\Sd\Bc]$.
\end{theorem}
\begin{proof}
    By definition, $[\tau(\Cons)\Inf\Sd\Bc] \subseteq [\Inf\Sd\Bc]$.
    Let $h$ be a $\Sd$ learner and $\La = \Inf\Sd\Bc(h)$.
    We use the $\Sd$-learner $g: D \mapsto \patch(h(D), D)$.
    \Cref{thm:patchMakesCons} yields that $g$ is globally consistent.

    Let $L \in \La$, $I \in \Inf(L)$ and $n \in \N$ with $W_{g(\content(I[n]))} = L$.
    Using \Cref{thm:patchMaintainsCorrect}, we have $g(\content(I[n])) \semequiv h(\content(I[n]))$ and thus $W_{g(\content(I[n]))} = L$, so $g$ is $\Bc$-learning $\La$.
\end{proof}

Note that \Cref{thm:plainglobalcons} is also a corollary of the far more intricate \Cref{thm:conswmonistrue}.
We can easily extend \Cref{thm:plainglobalcons} to all other interaction operators.

\begin{theorem}
    \label{thm:InfConsBcNoItAllEq}
    For all $\beta$ with $\Sd \trans \beta \trans \G$ we have $[\tau(\Cons)\Inf\beta\Bc] = [\Inf\G\Bc]$.
\end{theorem}
\begin{proof}
    One inclusion holds by definition.
    Using \Cref{thm:interactionForBc} and $\Sd \trans \beta$, we have $[\Inf\G\Bc] = [\Inf\Sd\Bc] = [\tau(\Cons)\Inf\Sd\Bc] \subseteq [\tau(\Cons)\Inf\beta\Bc]$.
\end{proof}

Using the approach to show $\CIt \semTransEq \Sd$ in \cite{kotzing2017normal}, a confluently iterative learner can fall back to a globally consistent set-driven learner, yielding the following result.

\begin{theorem}
    \label{thm:makeCItCons}
    We have $[\tau(\Cons)\Inf\CIt\Bc] = [\tau(\Cons)\Inf\Sd\Bc]$.
\end{theorem}

\begin{theorem}
    \label{thm:InfBcConsAllEq}
    For all interaction operators $\beta$ with $\CIt \trans \beta \trans \G$ we have $[\tau(\Cons)\Inf\beta\Bc] = [\Inf\G\Bc]$.
\end{theorem}
\begin{proof}
    This follows from \Cref{thm:InfConsBcNoItAllEq} and \Cref{thm:makeCItCons}.
\end{proof}

As an aside, this is in contrast to two observations for explanatory learning made by \textcite{aschenbach2018learning} that both $[\tau(\Cons)\Inf\G\Ex]$ and $[\InfCan\G\Cons\Ex]$ are proper subsets of $[\Inf\G\Cons\Ex]$.
Their proofs use the fact that global consistency can force a learner to make syntactic mindchanges that it otherwise would not have.

\begin{corollary}
    \label{thm:allConsEqualForPlainBc}
    We have $[\tau(\Cons)\Inf\G\Bc] = [\InfCan\G\Cons\Bc] = [\Inf\G\Bc]$.
\end{corollary}
\begin{proof}
    The statement $[\tau(\Cons)\Inf\G\Bc] = [\Inf\G\Bc]$ follows directly from \Cref{thm:InfBcConsAllEq}.
    For the other equality, one inclusion holds by definition and \Cref{thm:interactionForBc} yields $[\InfCan\G\Cons\Bc] \subseteq [\InfCan\G\Bc] = [\Inf\G\Bc]$.
\end{proof}

\section{Preserving monotonicity constraints}

We observe that our method of making learners consistent preserves some variants of monotonicity.
In particular, we patch the set of restrictions
$$\Delta_M = \set{\Mon, \DMon, \BMon, \SMon, \DSMon, \BSMon}.$$
For weak monotonicity, patching does not suffice, but we come up with another solution.

\begin{theorem}
    \label{thm:patchPreservesBDSMon}
    For all $\delta \in \Delta_M$, we have $[\tau(\Cons)\Inf\Sd\delta\Bc] = [\Inf\Sd\delta\Bc]$.
\end{theorem}
\begin{proof}
    One inclusion holds by definition.
    Let $h$ be a $\Inf\Sd\delta\Bc$-learner and \linebreak $\La = \Inf\Sd\delta\Bc(h)$.
    We use $g: D \mapsto \patch(h(D), D)$ as our modified learner.
    By \Cref{thm:patchMakesCons} and \Cref{thm:patchMaintainsCorrect}, $g$ is globally consistent and $\Bc$-learns $\La$.

    We proceed to show that the additions and subtractions of the $\patch$-function to $h$'s hypotheses do not violate $\delta$.
    Let $L \in \La$, $I \in \Inf(L)$ and $s,t \in \N$ with $s \leq t$.
    We abbreviate $S = \content(I[s])$ and $T = \content(I[t])$.

    \paragraph{Case 1: $\delta = \Mon$.}
    We use the fact that the patched-in elements continue to be patched-in and the patched-out elements are not considered in the definition of $\Mon$.
    Since $\neg(S) \cap L = \emptyset$ and $\pos(S) \subseteq L$, we get
    \begin{align*}
        W_{g(S)} \cap L &= W_{\patch(h(S), S)} \cap L \\
        &= ((W_{h(S)} \cup \pos(S)) \setminus \neg(S)) \cap L \\
        &= (W_{h(S)} \cup \pos(S)) \cap L \\
        &= (W_{h(S)} \cap L) \cup \pos(S).
    \end{align*}
    The same holds for $T$.
    Since $h$ is $\Mon$, we know $W_{h(S)} \cap L \subseteq W_{h(T)} \cap L$.
    Using $\pos(S) \subseteq \pos(T)$, we have
    $$W_{g(S)} \cap L = (W_{h(S)} \cap L) \cup \pos(S) \subseteq (W_{h(T)} \cap L) \cup \pos(T) = W_{g(T)} \cap L.$$

    \paragraph{Case 2: $\delta = \DMon$.}
    For this case, the idea is similar, but we need to do slightly more work.
    We use De Morgan's law and the fact that $L \cap \neg(S) = \emptyset$ to move the $L$ term inside.
    Since $\pos(S) \cup L = L$, we can then remove the $\pos(S)$ term.
    Finally, we use the two inclusions, one from $h$ being $\DMon$ and the other $\neg(S) \subseteq \neg(T)$.
    To get back, the same steps may be applied in reverse.
    \begin{align*}
        \overline{W_{g(S)}} \cap \overline L &= \overline{(W_{h(S)} \cup \pos(S)) \setminus \neg(S)} \cap \overline L \\
        &= \overline{((W_{h(S)} \cup \pos(S)) \setminus \neg(S)) \cup L} \\
        &= \overline{(W_{h(S)} \cup L) \setminus \neg(S)} \\
        &= (\overline{W_{h(S)}} \cap \overline L) \cup \neg(S) \\
        &\subseteq (\overline{W_{h(T)}} \cap \overline L) \cup \neg(T) \\
        &= \overline{W_{g(T)}} \cap \overline L.
    \end{align*}

    \paragraph{Case 3: $\delta = \BMon$.}
    Both restrictions are preserved as proven above.

    \paragraph{Case 4: $\delta = \SMon$.}
    For $\SMon$, the argument is similar to case 1.
    Still, elements that are patched-in stay patched-in.
    Since $h$ is strongly monotonic, we have for all $n \in \N$ that $W_{h(\content(I[n]))} \subseteq L$.
    Given that for any $n \in \N$, $\neg(I[n])$ and $L$ are disjoint, there are never any elements to patch out of the hypotheses of $h$.
    Formally, we have
    $$W_{g(S)} = W_{h(S)} \cup \pos(S) \subseteq W_{h(T)} \cup \pos(T) = W_{g(T)}.$$

    \paragraph{Case 5: $\delta = \DSMon$.}
    This is analogous to the case $\delta = \SMon$, with only the modification that now the added hypotheses are not considered and that $\neg(S)~\subseteq~\neg(T)$.
    Furthermore, since $h$ is dual strongly monotonic and identifies $L$ at some point, we have for all $n$ that $L \subseteq W_{h(\content(I[n]))}$.
    This means, that our patching never adds any more elements to the hypothesis, as they are all in already.
    We have
    $$W_{g(S)} = W_{h(S)} \setminus \neg(S) \supseteq W_{h(T)} \setminus \neg(T) = W_{g(T)}.$$

    \paragraph{Case 6: $\delta = \BSMon$.}
    Both restrictions are preserved as proven above.
\end{proof}

\begin{corollary}
    For all interaction operators $\Sd \trans \beta \trans \G$ and $\delta \in \Delta_M$, we have $[\tau(\Cons)\Inf\beta\delta\Bc] = [\InfCan\G\delta\Bc]$.
\end{corollary}
\begin{proof}
    One inclusion is trivial.
    Since $\delta\Bc$ is delayable, we can use \Cref{thm:interactionForDelayable}.
    With \Cref{thm:patchPreservesBDSMon}, we get
    $$[\InfCan\G\delta\Bc] = [\Inf\Sd\delta\Bc] = [\tau(\Cons)\Inf\Sd\delta\Bc] \subseteq [\tau(\Cons)\Inf\beta\delta\Bc].$$
\end{proof}

Finally, we also show that we can provide consistent versions of weakly monotonic learners.
In \Cref{thm:conswmonistrue} we have already seen that we consistency is no restriction for weakly monotonic learners, because neither poses a restriction at all.
Making dual weakly monotonic learners consistent is more difficult than the other variants of monotonicity.
Instead of just patching in the seen data, we use a poisoning approach.

\begin{theorem}
    \label{thm:makedualwmoncons}
    We have $[\tau(\Cons)\Inf\G\DWMon\Bc] = [\Inf\G\DWMon\Bc]$.
\end{theorem}
\begin{proof}
    We know by \Cref{thm:totalInteractionForDelayable} that $[\Inf\G\DWMon\Bc] = [\totalCp\Inf\Sd\DWMon\Bc]$.
    Let $h \in \totalCp$ be a $\Sd$-learner and $\La = \totalCp\Inf\Sd\DWMon\Bc(h)$.
    Consider the $\G$-learner $g$ that maps a finite sequence $\sigma \in \Seq(\N \times \bset)$ to the hypothesis
    $$W_{g(\sigma)} = \begin{cases}
        \pos(\sigma),              & \text{if } \exists \tau \subSeq \sigma: \pos(\tau) = \pos(\sigma) \land \pos(\sigma) \not\subseteq W_{h(\content(\tau))}, \\
        W_{h(\content(\sigma))},   & \text{else if } \Cons(h(\content(\sigma)), \sigma),                                                                                     \\
        \N \setminus \neg(\sigma), & \text{otherwise}.
    \end{cases}$$
    Intuitively, we use $h$'s hypotheses, but poison them if we can prove that they are wrong.
    This means that, if we observe that they are inconsistent with our positive information, we conjecture exactly this positive data, because our hypothesis then becomes inconsistent when new positive information is presented.
    We must then stick to this poisoned conjecture as long as no new positive information is shown.
    If $h$ includes wrong elements, we blow up our hypothesis to include everything but the negative data.
    This hypothesis becomes inconsistent once new negative information is presented.
    Notably, both poisoned hypotheses are already correct if no new positive or negative information is shown, respectively.

    We first show that $g$ is actually computable.
    To enumerate $W_{g(\sigma)}$, the program $g(\sigma)$ does the following for some input $x \in \N$:
    Firstly, if $x \in \pos(\sigma)$, it returns, if $x \in \neg(\sigma)$, it diverges.
    Then, it verifies that $\pos(\sigma)$ is included in $h$'s hypothesis for each $\tau \subSeq \sigma$, diverging if this is not the case.
    Finally, it tries to find any $y~\in~\set{x}~\cup~\neg(\sigma)$ that is also in $W_{h(\content(\sigma))}$, diverging if there is none.
    Hence, $g$ is computable by the s-m-n theorem.

    First, we show that $g$ $\Bc$-learns $\La$.
    Let $L \in \La$ and $I \in \Inf(L)$.
    Let $n_0 \in \N$ such that for all $n \geq n_0$ we have $W_{h(\content(I[n]))} = L$.
    This implies that \linebreak $\Cons(h(\content(I[n])), I[n])$ and hence $W_{g(I[n])}$ is either $\pos(I[n])$ or \linebreak $W_{h(\content(I[n]))}$.
    If $L$ is finite, then there is $n_1 \geq n_0$ such that $\pos(I[n_1]) = L$, so for all $n \geq n_1$ we have $W_{g(I[n])} = L$.
    If $L$ is infinite, let $n_2 > n_0$ be minimal such that $\pos(I[n_0]) \subsetneq \pos(I[n_2])$.
    Then, $W_{g(I[n_2])} = W_{h(\content(I[n_2]))} = L$ and we get by induction over all $n \geq n_2$ that $W_{g(I[n])} = L$.

    Clearly, $g$ is globally consistent.
    We proceed to show that $g$ is dual weakly monotonic.
    Let $s, t \in \N$ with $s < t$.
    We abbreviate $S = I[s]$ and $T = I[t]$.
    Suppose $\Cons(g(S), T)$.

    \paragraph{Case 1: $W_{g(S)} = \pos(S) \neq W_{h(\content(S))}$.}
    Therefore, there is $\tau \subSeq S$ with \linebreak $\pos(\tau) = \pos(S)$ and $\pos(S) \not\subseteq W_{h(\content(\tau))}$.
    Since $\Cons(g(S), T)$, we have \linebreak $\pos(T) \subseteq W_{g(S)} = \pos(S)$, so $\pos(S) = \pos(T)$.
    As $\tau \subSeq S \subSeq T$, we have \linebreak $W_{g(T)} = \pos(T) = W_{g(S)}$.

    \paragraph{Case 2: $W_{g(S)} = W_{h(\content(S))}$.}
    If $W_{g(T)} = \pos(T)$, then the assumption \linebreak $\Cons(g(S), T)$ gives us $W_{g(T)} = \pos(T) \subseteq W_{g(S)}$.
    Suppose $W_{g(T)} \neq \pos(T)$.
    The precondition $g(S) \semequiv h(\content(S))$ implies $\Cons(h(\content(S)), T)$.
    As $h$ is dual weakly monotonic, we have $W_{h(\content(S))} \supseteq W_{h(\content(T))}$.
    Furthermore, we have $W_{h(\content(S))} \cap \neg(T) = \emptyset$ and thus $W_{h(\content(T))} \cap \neg(T) = \emptyset$.
    As $W_{g(T)} \neq \pos(T)$, we have $\pos(T) \subseteq W_{h(\content(T))}$ and thus $\Cons(h(\content(T)), T)$.
    Therefore, $W_{g(T)} = W_{h(\content(T))} \subseteq W_{h(\content(S))} = W_{g(S)}$.

    \paragraph{Case 3: $W_{g(S)} = \N \setminus \neg(S)$.}
    Since $g$ is consistent and $\neg(S) \subseteq \neg(T)$, we have $W_{g(T)} \cap \neg(S) = \emptyset$ and thus $W_{g(T)} \subseteq \N \setminus \neg(S) = W_{g(S)}$.
\end{proof}

%% file: conclusions/further_research.tex
It is still unclear where dual weak monotonicity and combined weak monotonicity, also known as semantic conservativeness, should be located in our map.
Making a learner dual weakly monotonic appears to require a complete consistency check with previous hypotheses before enumerating elements.
This is why we do not believe that a modification similar to the one for classic weak monotonicity of learners can be achieved.
The separation of $\DWMon$ would also separate $\BWMon$ from $\True$.

\begin{conjecture}
    We have $[\Inf\G\DWMon\Bc] \subsetneq [\Inf\G\Bc]$.
\end{conjecture}

For $\BWMon$, we still think that the pattern of adding consistency holds.
We have observed that poisoning approaches work for adding consistency to weakly monotonic\footnote{not included, as it is implied by \Cref{thm:conswmonistrue}} and dual weakly monotonic learners.
To preserve combined monotonicity, it should be sufficient to add conditions to check whether either of the poisoned hypotheses has been given before.

\begin{conjecture}
    We have $[\Inf\G\BWMon\Bc] = [\tau(\Cons)\Inf\G\BWMon\Bc]$.
\end{conjecture}

Although it appears that consistency does not further restrict any of the monotonicity constraints and we know that the implications are not as strict as for $\Ex$ (see \Cref{thm:allConsEqualForPlainBc}), it is unclear whether an additional requirement for consistency narrows the learning power of learners under any other common restriction.

The restriction $\SemWb$ (semantically witness-based, \cite{kotzing2017normal}) could reveal more about the relationship of monotonic and cautious learners, because it is designed as a common lower bound of the two.
While we showed that all three reduced variants of cautiousness are properly restricting learners, we do not know how they relate to each other.
Furthermore, we do not know how the variants of cautiousness relate to other restrictions.
In particular, there may be an interesting relation to the monotonicity restrictions we mapped out in this work, as they all limit subset relations in the hypothesis sequence.

For semantic learning in general, there are several research directions.
There is very recent work by \textcite{marten2022isomorphisms} that investigates semantic learning restrictions in more abstract settings such as learning functions instead of languages.
It would also be interesting to see if stronger normal forms can be found, similar to what has been done by \textcite{kotzing2017normal} and \textcite{doskovc2021normal} for text-learning.

Lastly, there are of course more learning restrictions that are still missing from our map.
For example, the map for explanatory learning provided by \textcite{aschenbach2018learning} includes the semantic restrictions $\Dec$ (decisive) and $\NU$ (non-U-shaped).

%% file: references/references.bib
@inproceedings{kotzing2017normal,
  title={Normal forms in semantic language identification},
  author={K{\"o}tzing, Timo and Schirneck, Martin and Seidel, Karen},
  booktitle={International Conference on Algorithmic Learning Theory},
  pages={493--516},
  year={2017},
}

@inproceedings{kotzing2016towards,
  title={Towards an atlas of computational learning theory},
  author={K{\"o}tzing, Timo and Schirneck, Martin},
  booktitle={Symposium on Theoretical Aspects of Computer Science},
  year={2016},
}

@article{aschenbach2018learning,
  title={Learning from informant: relations between learning success criteria},
  author={Aschenbach, Martin and K{\"o}tzing, Timo and Seidel, Karen},
  journal={arXiv preprint},
  year={2018}
}

@article{lange1996monotonic,
  title={Monotonic and dual monotonic language learning},
  author={Lange, Steffen and Zeugmann, Thomas and Kapur, Shyam},
  journal={Theoretical Computer Science},
  volume={155},
  number={2},
  pages={365--410},
  year={1996},
  publisher={Elsevier}
}

@inproceedings{doskovc2021mapping,
  title={Mapping monotonic restrictions in inductive inference},
  author={Dosko{\v{c}}, Vanja and K{\"o}tzing, Timo},
  booktitle={Conference on Computability in Europe},
  pages={146--157},
  year={2021},
  organization={Springer}
}

@article{gold1967language,
  title={Language identification in the limit},
  author={Gold, E Mark},
  journal={Information and control},
  volume={10},
  number={5},
  pages={447--474},
  year={1967},
  publisher={Elsevier}
}

@inproceedings{doskovc2021normal,
  title={Normal forms for semantically witness-based learners in inductive inference},
  author={Dosko{\v{c}}, Vanja and K{\"o}tzing, Timo},
  booktitle={Conference on Computability in Europe},
  pages={158--168},
  year={2021},
  organization={Springer}
}

@article{lange1994characterization,
  title={Characterization of language learning front informant under various monotonicity constraints},
  author={Lange, Steffen and Zeugmann, Thomas},
  journal={Journal of Experimental \& Theoretical Artificial Intelligence},
  volume={6},
  number={1},
  pages={73--94},
  year={1994},
  publisher={Taylor \& Francis}
}

@article{kotzing2016map,
  title={A map of update constraints in inductive inference},
  author={K{\"o}tzing, Timo and Palenta, Raphaela},
  journal={Theoretical Computer Science},
  volume={650},
  pages={4--24},
  year={2016},
  publisher={Elsevier}
}

@article{blum1967machine,
  title={A machine-independent theory of the complexity of recursive functions},
  author={Blum, Manuel},
  journal={Journal of the ACM},
  volume={14},
  number={2},
  pages={322--336},
  year={1967},
  publisher={ACM New York, NY, USA}
}

@article{case1994infinitary,
  title={Infinitary self-reference in learning theory},
  author={Case, John},
  journal={Journal of Experimental \& Theoretical Artificial Intelligence},
  volume={6},
  number={1},
  pages={3--16},
  year={1994},
  publisher={Taylor \& Francis}
}

@phdthesis{seidel2021modelling,
  title={Modelling binary classification with computability theory},
  author={Seidel, Karen},
  year={2021},
  school={Universit{\"a}t Potsdam}
}

@article{case1983comparison,
  title={Comparison of identification criteria for machine inductive inference},
  author={Case, John and Smith, Carl},
  journal={Theoretical Computer Science},
  volume={25},
  number={2},
  pages={193--220},
  year={1983},
  publisher={Elsevier}
}

@article{kotzing2017solution,
  title={A solution to Wiehagen’s thesis},
  author={K{\"o}tzing, Timo},
  journal={Theory of Computing Systems},
  volume={60},
  number={3},
  pages={498--520},
  year={2017},
  publisher={Springer}
}

@book{rogers1987theory,
  title={Theory of recursive functions and effective computability},
  author={Rogers Jr, Hartley},
  year={1987},
  publisher={MIT press}
}

@article{kleene1952introduction,
  title={Introduction to metamathematics},
  author={Kleene, Stephen Cole},
  year={1952}
}

@article{jantke1991monotonic,
  title={Monotonic and non-monotonic inductive inference},
  author={Jantke, Klaus P},
  journal={New Generation Computing},
  volume={8},
  number={4},
  pages={349--360},
  year={1991},
  publisher={Springer}
}

@inproceedings{wiehagen1990thesis,
  title={A thesis in inductive inference},
  author={Wiehagen, Rolf},
  booktitle={International Workshop on Nonmonotonic and Inductive Logic},
  pages={184--207},
  year={1990},
  organization={Springer}
}

@article{angluin1980inductive,
  title={Inductive inference of formal languages from positive data},
  author={Angluin, Dana},
  journal={Information and control},
  volume={45},
  number={2},
  pages={117--135},
  year={1980},
  publisher={Elsevier}
}

@article{osherson1982learning,
  title={Learning strategies},
  author={Osherson, Daniel N and Stob, Michael and Weinstein, Scott},
  journal={Information and Control},
  volume={53},
  number={1-2},
  pages={32--51},
  year={1982},
  publisher={Academic Press}
}

@thesis{marten2022isomorphisms,
    title={Isomorphisms and embeddings between limit learning settings},
    author={Marten, Paula},
    type={Bachelor's thesis},
    year={2022},
    school={Universit{\"a}t Potsdam}
}

@book{wexler1980formal,
  title={Formal principles of language acquisition},
  author={Wexler, Kenneth and Culicover, Peter W},
  year={1980},
  publisher={MIT Press (MA)}
}

@phdthesis{schafer1984eingabeabhangigkeit,
  title={{\"U}ber Eingabeabh{\"a}ngigkeit und Komplexit{\"a}t von Inferenzstrategien},
  author={Sch{\"a}fer-Richter, Gisela},
  year={1984},
  school={RWTH Aachen}
}

@inproceedings{case2010strongly,
  title={Strongly non-u-shaped learning results by general techniques},
  author={Case, John and K{\"o}tzing, Timo},
  booktitle={COLT},
  volume={2010},
  pages={181--193},
  year={2010},
  organization={Citeseer}
}


%% file: references/strings.bib
@Preamble
{
    {
    \newcommand{\bibciac}[2]{Proceedings of the #1 Conference on Algorithms and Complexity (CIAC'#2)}
    \newcommand{\bibdac}[2]{Proceedings of the #1 Annual Design Automation Conference (DAC'#2)}
    \newcommand{\bibinvisau}[1]{Proceedings of the Australian Symposium on Information Visualisation (invis.au #1)}
    \newcommand{\bibieeepdp}[2]{Proceedings of the #1 IEEE Symposium on Parallel and Distributed Processing #2}
    \newcommand{\bibieeecs}[1]{Proceedings of the IEEE International Symposium on Circuits and Systems #1}
    \newcommand{\bibcccg}[2]{Proceedings of the #1 Canadian Conference on Computational Geometry (CCCG'#2)}
    \newcommand{\bibswat}[2]{Proceedings of the #1 Scandinavian Workshop on Algorithm Theory (SWAT'#2)}
    \newcommand{\bibipco}[2]{Proceedings of the #1 International Conference on Integer Programming and Combinatorial Optimization (IPCO'#2)}
    \newcommand{\bibsofsem}[2]{Proceedings of the #1 Conference on Current Trends in Theory and Practice of Computer Science (SOFSEM'#2)}
    \newcommand{\bibstoc}[2]{Proceedings of the #1 Annual ACM Symposium on Theory of Computing (STOC'#2)}
    \newcommand{\bibfocs}[2]{Proceedings of the #1 Annual Symposium on Foundations of Computer Science (FOCS'#2)}
    \newcommand{\bibsoda}[2]{Proceedings of the #1 Annual ACM-SIAM Symposium on Discrete Algorithms (SODA'#2)}
    \newcommand{\bibgd}[2]{Proceedings of the #1 International Symposium on Graph Drawing (GD'#2)}
    \newcommand{\bibinfovis}[1]{Proceedings of the IEEE Symposium on Information Visualization (InfoVis'#1)}
    \newcommand{\bibvis}[1]{Proceedings of the IEEE Conference on Visualization (Vis'#1)}
    \newcommand{\bibpvis}[1]{Proceedings of the IEEE Pacific Visualisation Symposium (PacificVis'#1)}
    \newcommand{\bibsoftvis}[2]{Proceedings of the #1 ACM Symposium on Software Visualization (SoftVis'#2)}
    \newcommand{\bibeurocg}[2]{Proceedings of the #1 European Workshop on Computational Geometry (EuroCG'#2)}
    \newcommand{\bibsocg}[2]{Proceedings of the #1 Annual Symposium on Computational Geometry (SoCG'#2)}
    \newcommand{\bibwads}[2]{Proceedings of the #1 International Symposium on Algorithms and Data Structures (WADS'#2)}
    \newcommand{\bibwg}[2]{Proceedings of the #1 Workshop on Graph-Theoretic Concepts in Computer Science (WG'#2)}
    \newcommand{\bibgta}{Proceedings of the Conference at Graph Theory and Applications}
    \newcommand{\bibisaac}[2]{Proceedings of the #1 International Symposium on Algorithms and Computation (ISAAC'#2)}
    \newcommand{\bibcocoon}[2]{Proceedings of the #1 Annual International Conference on Computing and Combinatorics (COCOON'#2)}
    \newcommand{\bibtamc}[2]{Proceedings of the #1 Annual Conference on Theory and Applications of Models of Computation (TAMC'#2)}
    \newcommand{\bibicalp}[2]{Proceedings of the #1 International Colloquium on Automata, Languages and Programming (ICALP'#2)}
    \newcommand{\biblatin}[2]{Proceedings of the #1 Latin American Symposium (LATIN'#2)}
    \newcommand{\bibesa}[2]{Proceedings of the #1 Annual European Symposium on Algorithms (ESA'#2)}
    }
}

@String{Elsevier
= {Elsevier Science Publishers}}

@String{ACM
= {ACM Press}}
